\documentclass[fleqn,usenatbib]{mnras}

\usepackage[T1]{fontenc}
\usepackage{ae,aecompl}

\usepackage{graphicx}        
\usepackage{amsmath}         
\usepackage{amssymb}         

\usepackage{booktabs}        
\usepackage{scalerel}        
\usepackage{xcolor,listings} 
\usepackage{upquote}         
\usepackage{fixmath}         
\usepackage{multirow}        

\usepackage{xcolor}

\newcommand*{\dirplot}{plot/png_300}
\newcommand*{\ext}{png}

\newcommand*{\Sun}{\protect\scalebox{0.7}{$\odot$}}

\newcommand*{\Range}[2]{[#1\,\text{-}\,#2]}

\newcommand*{\pd}[1]{\!\times\!10^{#1}}

\newcommand*{\units}[1]{\scalebox{0.8}{(#1)}}

\newcommand*{\SM}[1]{{\scaleto{\rm #1}{3.5pt}}}

\newcommand*{\var}[1]{\mbox{\footnotesize$(#1)$}}

\newcommand*{\minus}{\scalebox{0.75}[1.0]{$-$}}
\newcommand*{\plus}{\scalebox{0.75}[1.0]{$+$}}

\newcommand*{\Gband}{\mbox{$G\:\!\text{-band}$} }

\newcommand*{\BV}{B\textnormal{-}\:\:\!\!\!V}

\newcommand*{\BPRP}{G_{\rm BP}\!-\!G_{\rm RP}}

\newcommand*{\GtrSim}{\smallrel\gtrsim}
\newcommand*{\LessSim}{\smallrel\lesssim}

\newcommand*{\Approx}{\smallrel\sim}

\makeatletter
\newcommand*{\smallrel}[2][.8]{%
  \mathrel{\mathpalette{\smallrel@{#1}}{#2}}%
}
\newcommand*{\smallrel@}[3]{%
  \sbox0{$#2\vcenter{}$}%
  \dimen@=\ht0 %
  \raise\dimen@\hbox{%
    \scalebox{#1}{%
      \raise-\dimen@\hbox{$#2#3\m@th$}%
    }%
  }%
}

\definecolor{mymauve}{rgb}{0.58,0,0.82}
\definecolor{mygreen}{rgb}{0,0.6,0}

\title[Tidal stream of NGC 3201]{The tidal stream generated by the globular cluster NGC 3201}
\author[C. G. Palau, J. Miralda-Escud\'e]{
Carles G. Palau,$^{1}$\thanks{E-mail: cgarcia@icc.ub.edu}
Jordi Miralda-Escud\'e,$^{1,2}$\thanks{E-mail: miralda@icc.ub.edu}
\\
$^{1}$Institut de Ci\`encies del Cosmos, Universitat de Barcelona (UB-IEEC), Mart\'\i\ i Franqu\`es 1, E-08028 Barcelona, Catalonia, Spain.\\
$^{2}$Instituci\'o Catalana de Recerca i Estudis Avan\c cats, E-08028 Barcelona, Catalonia, Spain.\\
}

\date{Accepted XXX. Received YYY; in original form ZZZ}

\pubyear{2020}

\begin{document}
\label{firstpage}
\pagerange{\pageref{firstpage}--\pageref{lastpage}}
\maketitle

\begin{abstract}
We detect a tidal stream generated by the globular cluster NGC 3201 extending
over $\Approx140$ degrees on the sky, using the \textit{Gaia} DR2 data, with the maximum-likelihood method we presented previously to study the M68 tidal stream. Most
of the detected stream is the trailing one, which stretches in the southern
Galactic hemisphere and passes within a close distance of 3.2 kpc from the Sun,
therefore making the stream highly favourable for discovering relatively
bright member stars, while the leading arm is further from us and behind a disc
foreground that is harder to separate from. The cluster has just crossed the
Galactic disc and is now in the northern Galactic hemisphere, moderately
obscured by dust, and the part of the trailing tail closest to the cluster is
highly obscured behind the plane. We obtain a best-fitting model of the stream
which is consistent with the measured proper motion, radial velocity, and
distance to NGC 3201, and show it to be the same as the previously detected
Gj\"oll stream by Ibata et al. We identify $\Approx200$ stars with the highest
likelihood of being stream members using only their \textit{Gaia} kinematic data. Most
of these stars (170) are photometrically consistent with being members of
NGC 3201 when they are compared to the cluster H-R diagram, only once a
correction for dust absorption and reddening by the Galaxy is applied. The
remaining stars are consistent with being random foreground objects according
to simulated data sets. We list these 170 highly likely stream member stars.
\end{abstract}




\section{Introduction}

 Stellar streams associated with globular clusters are formed when a
cluster is tidally disrupted by its host galaxy. For a disc galaxy like the
Milky Way, stars are stripped from the cluster especially when it approaches
the centre of the Galaxy or crosses the Galactic disc. Each tidal shock
populates the leading and trailing arms of the tidal stream, with escaped
stars approximately following the orbit of the progenitor. This makes stellar
streams useful tools to constrain the Galactic potential by fitting models of the stellar stream orbits
to the observations, particularly if accurate observations of proper motions and radial velocities are available
for the stream member stars. So far, many stellar streams have been discovered,
\cite[e.g.][]{2016ASSL..420...87G,2018MNRAS.481.3442M,2018ApJ...862..114S,2019ApJ...884..174G}.
The publication of the \textit{Gaia} Data Release 2 (GDR2) provides a promising
opportunity to discover new ones and to use the measured motions to study the
Milky Way potential.

 Several papers have developed methods to detect stellar streams in star
catalogues with data on photometry, proper motions or radial velocities
\citep[e.g.][]{2015ApJ...801...98S,2017MNRAS.469..721M,2018MNRAS.477.4063M}. In a previous paper
\citep[][hereafter PM19]{2019MNRAS.488.1535P}, we developed a new statistical method based on
maximum-likelihood analysis designed to detect stellar streams associated
with a known stellar system such as a globular cluster, when a
small number of stream members appear superposed on a large catalogue of
foreground stars. The method searches for a statistically significant
overdensity of stars compared to a phase-space density model of the Milky
Way. A stream model is constructed with free parameters that include
the potential model of the Galaxy determining the orbits, plus the distance
and velocity of the globular cluster within the constraints of the available
observations. Numerical simulations of the stream are used to construct its
phase-space density model. Then, the likelihood of each star in a catalogue is
computed for the simulated model of the stream, given the observed phase-space
coordinates and their observational errors. The model-free parameters that
maximize the likelihood function are obtained, and a statistical test for this
best-fitting model is performed to infer whether the stream exists or not. If the
statistical evidence for the stream existence is sufficient, we use the stream
density model to select stars that are most likely to be stream members based
on the kinematic evidence. Finally, our final selection is obtained by
requiring the stars to be also compatible with the H-R diagram of the cluster,
assuming the distance to each star to be that predicted by the stream model.

 Applying this method to the globular cluster M68, we found a long tidal
stream stretching over the North Galactic hemisphere, and passing about 5 kpc
from the Sun. This stream was found to match the stellar stream named Fj\"orm,
independently discovered by \citet{2019ApJ...872..152I}. For that study,
absorption and reddening by Galactic dust was neglected when using the
photometric observations to require stream members to be compatible with the
H-R diagram of M68. We have further checked if other streams generated by globular clusters
can be found in the GDR2 catalogue. Here, we study the case of NGC 3201, for
which we also find a new stellar stream in which the effect of dust absorption
and reddening is large and crucial for recognizing the stream members.

In Section \ref{sec2} we describe NGC 3201 and our simulation of its tidal
stream, and discuss the expected background using a simulation of the
\textit{Gaia} catalogue. In Section \ref{sec3}, our statistical method is applied to
select the GDR2 star candidate members of the NGC 3201 tidal stream and to
estimate its statistical significance, and we conclude in Section \ref{sec4}.

\section{Simulations of the NGC 3201 tidal stream and detection method}
\label{sec2}

 Our stream detection method, fully described in PM19, starts by
computing an initial simulation of the tidal stream of NGC 3201 using a
fiducial model for the Galactic potential and the central observed
values of the velocity and distance to the globular cluster. Then, a
bundle of possible stream models is computed by considering a range of
parameter values for both the Galactic potential and the globular
cluster kinematics, which is used to pre-select a sample of stars in
GDR2 as possible candidates of the stream, greatly reducing the number
of stars to be used in the final model fit of the stream. One important
difference we will find in this work compared to our previous one on the
globular cluster M68 in PM19 is that NGC 3201 is close to the Galactic
plane, at $b=8.64$ deg, with a high density of foreground stars and dust
obscuration. We will start ignoring the presence of dust obscuration in
this section (like we did in PM19 for the M68 tidal stream), but in the
next section we shall include a model for obscuration and reddening,
showing how it has substantial impact in our final selection of candidate
members of the NGC 3201 tidal stream.

 The NGC 3201 cluster is $\sim$ 5 kpc away from the Sun, near the
Galactic plane and at longitude $l=277.23$ deg, and has an extreme
radial velocity of $494\, {\rm km}\,{\rm s}^{-1}$, the highest of all
globular clusters in the Milky Way, which indicates a retrograde orbit
coming from a large apocentre. The implied long orbital period motivates
searches for a tidal stream associated to this cluster, which may have
formed from its outer envelope and not have been exposed to a large
degree of phase mixing during its orbital history. Some evidence for
this tidal stream has been pointed out in \citet{2010ApJ...721.1790C},
who noted aligned star clumps of 2MASS sources in the cluster envelope.
\citet{2014AandA...572A..30K} obtained similar conclusions using stars
with radial velocity from the RAVE survey, identifying unbound stars
extending a few arc minutes away from the cluster. This was extended by
\citet{2016MNRAS.457.2078A}, who found tidal stream candidates out to
$\sim 80$ deg from the cluster. Recent work using \textit{Gaia} data has
confirmed these observations, reporting an excess of RR Lyrae
\citep{2019MNRAS.483.1737K} and a high velocity dispersion profile
beyond the Jacobi radius together with aligned stellar overdensities
near the cluster \citep{2019ApJ...887L..12B}.

\subsection{Initial stream simulation}

 We carry out a fiducial simulation of the formation and evolution of
the tidal stream of NGC 3201, following the method that is described in
detail in section 2.5 of PM19. Briefly, the method consists of
integrating the orbits of tests particles initially distributed in a
fixed Plummer potential that models the globular cluster, which is at
the same time orbiting in a fixed potential of the Milky Way. Initial
conditions for the cluster orbit are taken from 
\citet{1996AJ....112.1487H,2010arXiv1012.3224H} for the heliocentric
distance $r_{\rm h}$, right ascension $\alpha$, and declination $\delta$,
and radial velocity $v_r$, and we use the proper motion
$\mu_{\alpha *}=\mu_\alpha\cos\var{\delta}$ and $\mu_{\delta}$ from the
\textit{Gaia} catalogue \citep{2018AA...616A..12G}. Central observed values and
errors are listed in Table \ref{NGC3201_DH}. Note that the heliocentric
distance was obtained from modelling the H-R diagram, with an estimated
error of 2.3 per cent, because the parallax error from \textit{Gaia} is
much larger.

\begin{table}
\caption[]{\small{Mass, core radius, present position, radial velocity,
and proper motion of the NGC 3201 globular cluster. Dark halo mass
density profile parameters used for tidal stream simulation, and
computed cluster orbit properties.}}
\begin{center}
\begin{tabular}{lllc}
\toprule
\multicolumn{3}{l}{\textbf{Properties NGC 3201}}&\textbf{Ref.}\\
\midrule
$M_{\rm gc}$&\units{M$_{\odot}$}&$(6.47\pm0.45)\pd{4}$&[1]\\
$a_{\rm gc}$&\units{pc}&$4.9$&[1]\\
\midrule
$r_{\rm h}$&\units{kpc}&$4.9\pm0.11$&[2]\\
$\delta$&\units{deg}&$-46.4125$&[3]\\
$\alpha$&\units{deg}&$154.3987$&[3]\\
$v_r$&\units{km s$^{-1}$}&$494\pm0.2$&[2]\\
$\mu_\delta$&\units{mas yr$^{-1}$}&$-1.9895\pm0.002$&[3]\\
$\mu_\alpha$&\units{mas yr$^{-1}$}&$12.0883\pm0.0031$&[3]\\
\midrule
\multicolumn{3}{l}{\textbf{Dark halo properties}}&\\
\midrule
$\rho_{0 \rm dh}$&\units{M$_{\odot}$ kpc$^{-3}$}&$7.27\pd{6}$&[4]\\
$a_{1 \rm dh}$&\units{kpc}&$18.59$&[4]\\
$a_{3 \rm dh}$&\units{kpc}&$16.17$&[4]\\
$\alpha_{\rm dh}$&-&$1$&[4]\\
$\beta_{\rm dh}$&-&$3.102$&[4]\\
\midrule
$q$&-&$0.87$\\
$q_{\varPhi}$&-&$0.94$\\
$c_{200}$&-&$10.4$\\
$r_{200}$&\units{kpc}&$175.9$\\
$M_{200}$&\units{M$_{\odot}$}&$6.37\pd{11}$\\
\midrule
\multicolumn{3}{l}{\textbf{Orbit properties}}&\\
\midrule
$r_{\rm peri}$&\units{kpc}&$7.71$&\\
$r_{\rm apo}$&\units{kpc}&$43.25$&\\
$L_{z}$&\units{km s$^{-1}$ kpc}&$2765.45$&\\
\bottomrule
\end{tabular}
\end{center}

\begin{tabular}{l}
\text{[1]}: \citet{2017MNRAS.471.3668S}\\
\text{[2]}: \citet{1996AJ....112.1487H,2010arXiv1012.3224H}\\
\text{[3]}: \citet{2018AA...616A..12G}\\
\text{[4]}: PM19\\
\end{tabular}

\label{NGC3201_DH}
\end{table}

 The cluster orbit is integrated first as that of a test particle in the
Milky Way potential model described in PM19. This model has fixed bulge
and disc components, and an axisymmetric oblate dark halo with the
parameters listed in Table \ref{NGC3201_DH}, obtained as the best fit of
the M68 tidal stream and other observational constraints in PM19. The
parameters are a constant of proportionality $\rho_{0 \rm dh}$, a
Galactic plane scale length $a_{1 \rm dh}$, a vertical scale height
$a_{3 \rm dh}$, and inner slope $\alpha_{\rm dh}$ and an outer slope
$\beta_{\rm dh}$. This profile is very close to an oblate NFW profile
(which has $\beta_{\rm dh}=3$; \citet{1996ApJ...462..563N}), with a concentration
parameter $c=10.4$, virial radius $r_{200} = 175.9\, {\rm kpc}$,
a total mass $M_{200}=6.37\pd{11}\, {\rm M_{\Sun}}$, and an axis ratio
$q=a_{3 \rm dh}/a_{1 \rm dh}=0.87$. The corresponding potential
flattening is $q_\Phi=0.94$ at the cluster position.

\begin{figure*}
\begin{tabular}{c}
\includegraphics[]{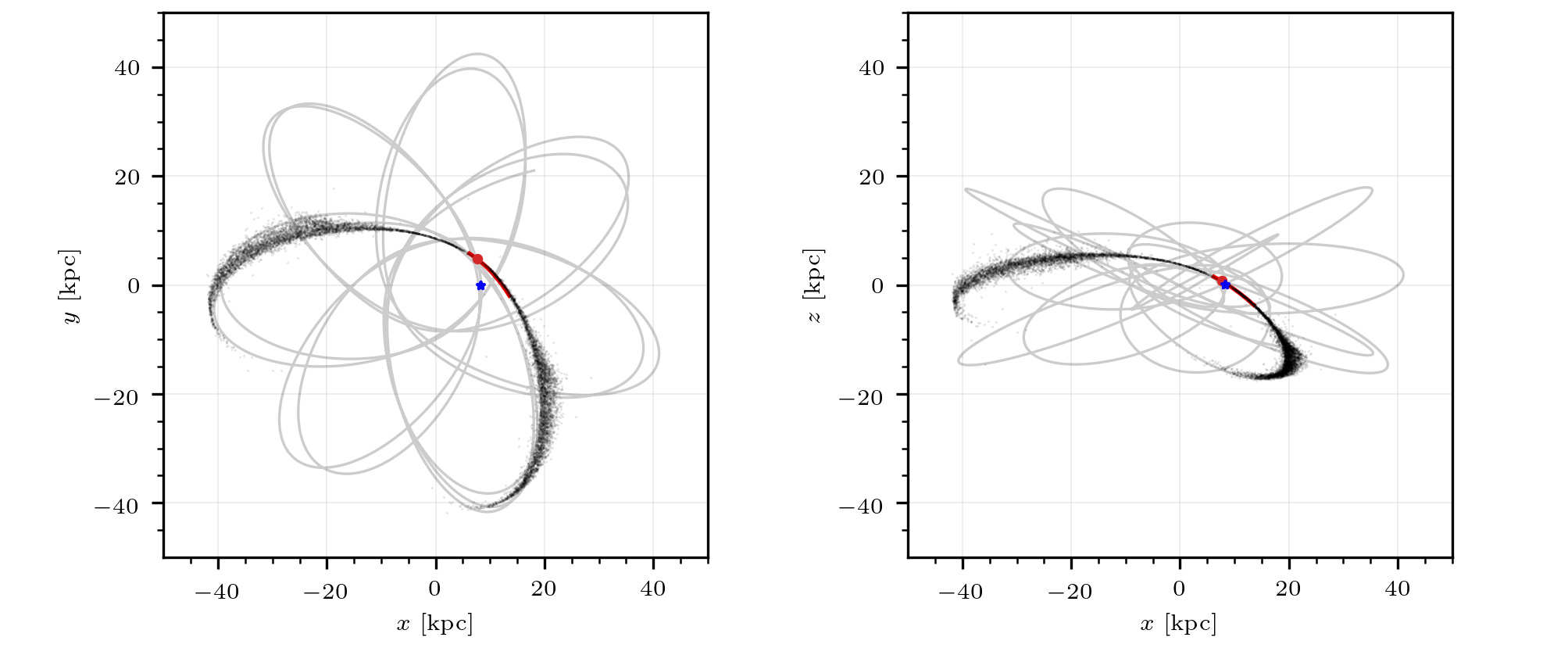}\\
\includegraphics[]{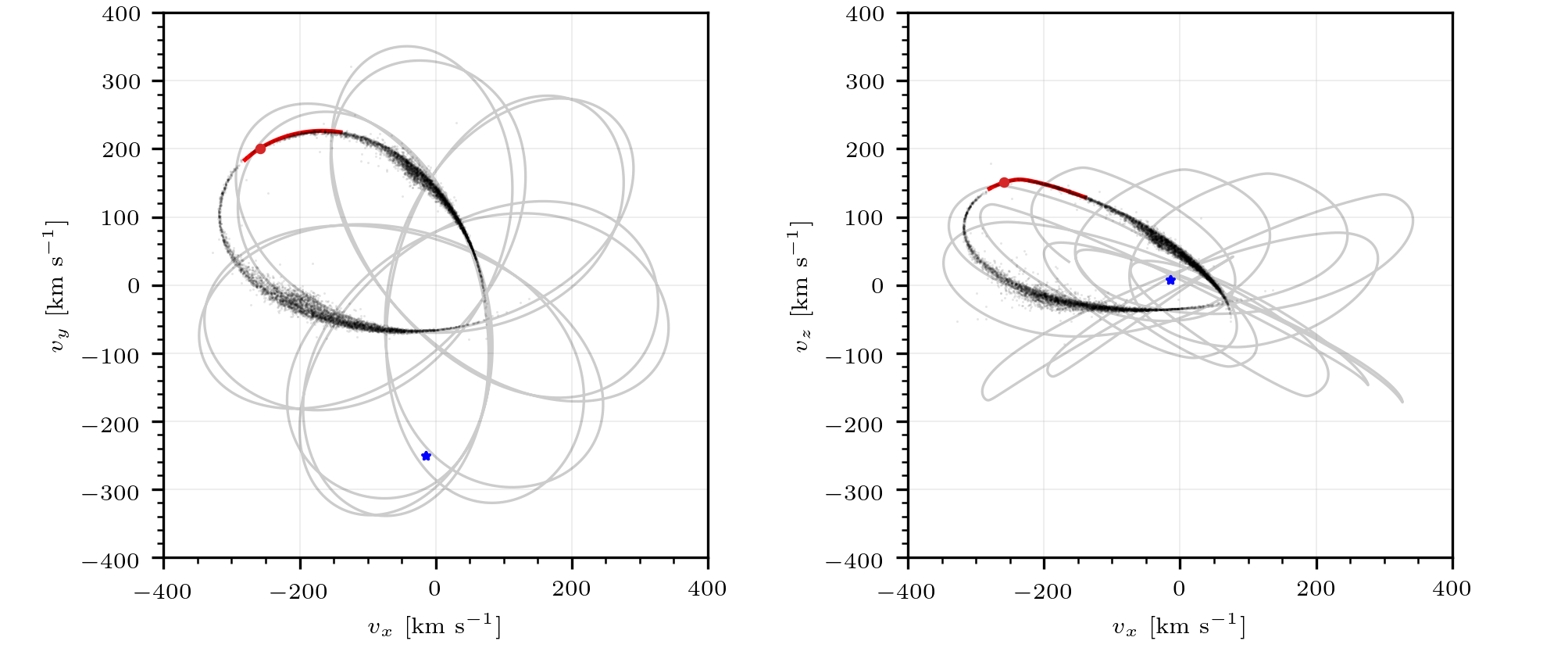}
\end{tabular}
\caption{Computed orbit of NGC 3201 over the last 10 Gyr (grey line). The section from $-60\, {\rm Myr}$ (trailing arm) to $+10\, {\rm Myr}$ (leading arm) from the present position of the cluster (red dot) is highlighted in red. The black dots indicate the position of $10^4$ tidal stream stars that have escaped the cluster potential, and the blue star is the current position of the Sun. \textit{Top:} Projection on the Galactic disc plane ($x,y$) and the ($x,z$) plane. \textit{Bottom:} Same projections in velocity space.}
\label{stream_NGC3201}
\end{figure*}

 We plot the cluster orbit as a grey line in Galactocentric Cartesian
coordinates in the top panels of Figure \ref{stream_NGC3201}, on the x-y
and x-z projections. The current cluster position and the Sun are
marked as a red and blue dot, respectively. We highlight in red the section of the orbit from a time $-60\, {\rm Myr}$ (trailing arm) to $+10\, {\rm Myr}$ (leading arm) from the present time cluster position. The orbit in the $v_x$-$v_y$
and $v_x$-$v_z$ velocity space projections is shown in the bottom panels.
Our computed orbital pericentre and apocentre of NGC 3201 and its
vertical angular momentum component, listed in Table \ref{NGC3201_DH},
are similar to the orbit of the tidal stream designated as Gj\"oll,
discovered by \citet{2019ApJ...872..152I}, which has
$r_{\rm peri} = 7.96\pm0.22\, {\rm kpc}$, $r_{\rm apo} = 31.9\pm4.4\,
{\rm kpc}$, and $L_z = 2721\pm159\, {\rm km}\,{\rm s}^{-1}\,{\rm kpc}$.
We shall show in this paper that the Gj\"oll tidal stream does in fact
originate from the NGC 3201 globular cluster. The differences in the
orbital parameters are consistent with observational and modelling
uncertainties and the expected difference between the cluster orbit and
the tidal stream. 

We compute the orbits of $10^6$ tidal stream stars as mentioned earlier
and described in detail in PM19, using a fixed Plummer sphere model for
the globular cluster potential with a scale parameter $a_{\rm gc}=4.9\, {\rm pc}$ and a total stellar mass $M_{\rm gc} = (6.47\pm0.45)\pd{4}\, {\rm M}_{\Sun}$ from \citet{2017MNRAS.471.3668S}. In the same reference, the inferred dynamical mass for a King-Michie model is included being a factor 2 higher. An accurate estimate of the cluster mass is not relevant for our analysis since we are 
assuming a fixed mass throughout the evolution and the details of the phase-space distribution of the stream are not relevant to our detection method. The tidal stream orbits are started at the
cluster position 10 Gyr ago, and integrated forwards in time up to the
present, assuming the Plummer sphere potential follows the cluster orbit
previously computed as a test particle in the Milky Way potential, and
simply adding the Plummer and Milky Way potentials. Of all the simulated
stream stars that have escaped further than $0.1$ deg from the cluster
centre, a randomly selected subset of $10^4$ of them are shown in Figure
\ref{stream_NGC3201} as small black dots (we do not plot all of them
only to better visualize their distribution). As seen in these plots,
the cluster is on a relatively low inclination orbit and has recently
crossed the Galactic disc moving upwards. The tidal shock it experienced
may be the explanation for the overdensities observed in the cluster
neighbourhood. The part of the tidal stream closest to us is the
trailing arm, at $\sim 4$ kpc from us and 1 to 2 kpc below the disc. The
large population of stars at the ends of the tidal stream is due to our
initial conditions, which do not have any radial cut-off in the initial
distribution of stars in the Plummer sphere, so many stars escape during
the first orbits. In reality, the existence of any initial
overdensities from the time the globular cluster started tidally
interacting with the Milky Way depends on the history of the cluster and
the Milky Way potential, which are likely to cause phase mixing and
violent relaxation to a much greater extent than in our simple, fixed
potential simulation.

\begin{figure*}
\includegraphics[]{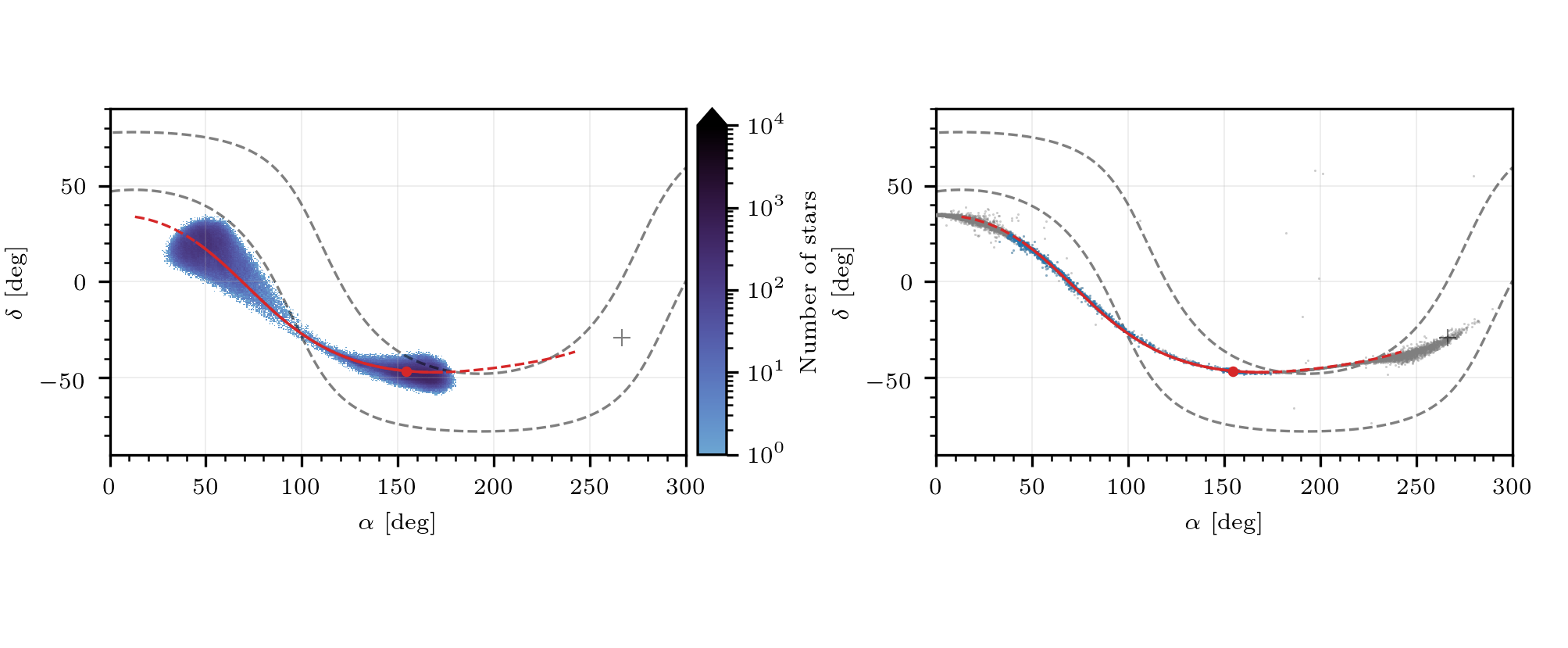}\\
\vspace{-4em}
\includegraphics[]{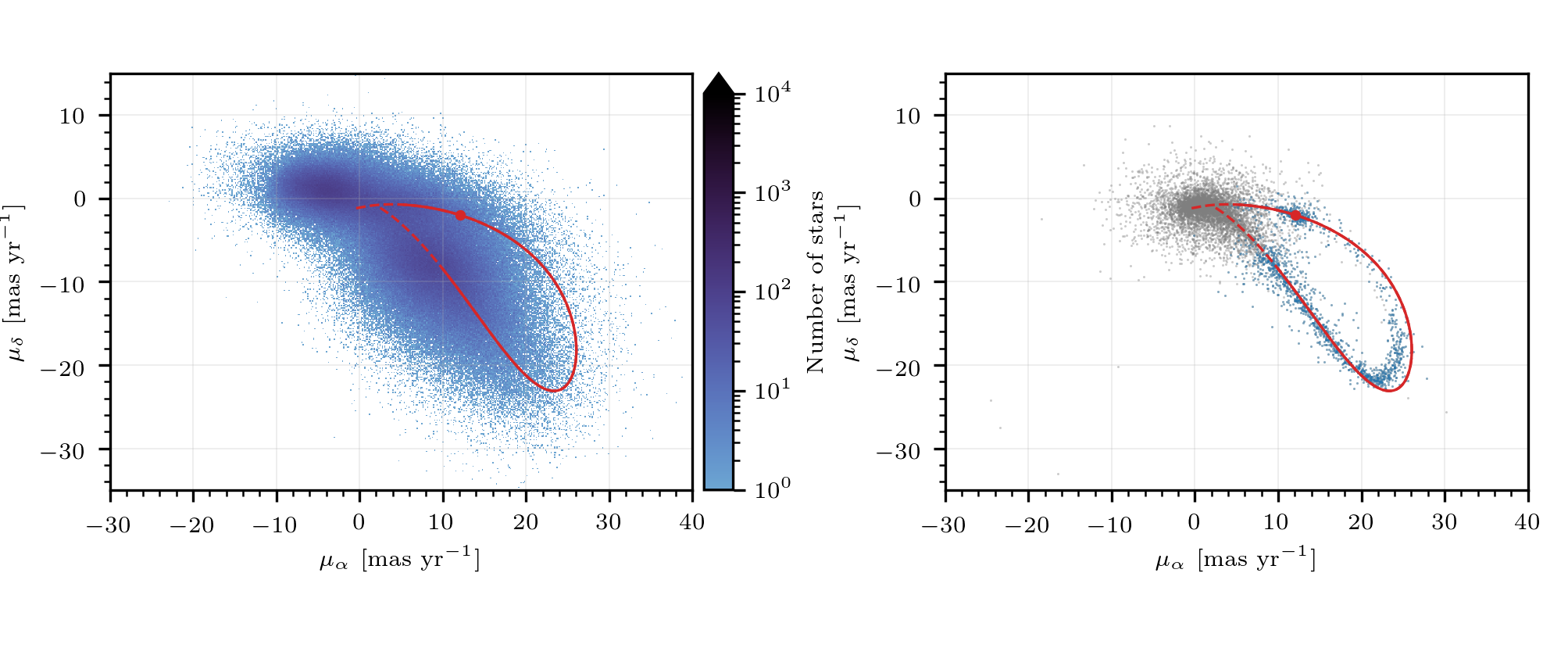}
\caption{Computed orbit of NGC 3201, with its present position shown as
the red dot, from $-60\, {\rm Myr}$ to $+10\, {\rm Myr}$ from the
present (solid red line), and from $-200\, {\rm Myr}$ to
$+200\, {\rm Myr}$ (dashed red line). The grey cross is the Galactic centre,
and grey dashed lines mark Galactic latitude $b=\pm15$ deg.
\textit{Left-hand panels:} Blue dots show coordinates and proper motions of
stars in the GOG18 catalogue pre-selected for our search for stream
candidates. \textit{Right-hand panels:} Coordinates and proper motions of the
simulated tidal stream stars are shown as blue dots
if they are in our pre-selected sample, and as grey dots if they are not.}
\label{gog18_pre_short}
\end{figure*}

  The cluster orbit is shown in equatorial coordinates in the top panels
of Figure \ref{gog18_pre_short}, from 200 Myr in the past to 200 Myr in
the future, as the dashed red line, with the red dot indicating the
present position. The orbital path from 60 Myr ago to 10 Myr in the
future is highlighted as the red solid line. We shall see that this is
the part of the tidal stream where stars are most easily identified from
proper motions in the \textit{Gaia} catalogue. The Galactic centre is indicated
by the grey cross, and dashed grey lines show the Galactic latitude
lines at $b\pm15$ deg. We also plot the cluster orbit in proper motion
space in the bottom panels of Figure \ref{gog18_pre_short}. The interval
that is highlighted as the solid line lies in a region of higher proper
motion than the rest of the orbit, which helps us to reduce the density of
foreground stars and facilitates the identification of stream
candidates.

\subsection{Tests with the simulated \textit{Gaia} catalogue}\label{tests}

 We now use the stars in our model tidal stream to simulate how they
would be observed with \textit{Gaia}. While computing proper motions
and parallaxes from the kinematics of each star in the tidal stream is
trivial, the observational errors depend on the magnitude and colour of
the stars, which we therefore need to simulate. We follow the same
procedure as in PM19: we first obtain the H-R diagram of NGC 3201 from
the \textit{Gaia} data itself, by selecting a total of 7064 stars that are within
0.14 deg of the globular cluster centre and pass additional
conditions specified in Appendix \ref{App0}. This H-R diagram is shown in Figure \ref{HR_NGC3201}, where the derived absolute magnitude without dust correction, $M_G'$, computed assuming a distance $r_h=4.9$ kpc, is plotted against the observed colour index $(\BPRP)'$, and the primes generally indicate that magnitudes are not corrected for dust extinction. We randomly assign to each escaped star an absolute magnitude and colour from this H-R diagram, and compute an
apparent magnitude using its simulated distance. Dust obscuration and
reddening is not taken into account here, this will be included only in
the next section when selecting stream candidates from the real data.

 We then generate measurement error covariance matrices for the
phase-space coordinates of the simulated stream stars using the Python
toolkit PYGAIA\footnote{\url{https://pypi.org/project/PyGaia/}}, and
following the same procedure as in section 3.2 of PM19.
We assign these errors to each star, and also use them to alter the
positions and velocities from the stream model by variations generated
as Gaussian distributions following the same error covariance matrices.
We plot as grey dots in the right-hand panels of Figure \ref{gog18_pre_short} the 8319 stars with \Gband magnitude $G<21$ that constitute our simulated catalogue of the stellar stream as seen by \textit{Gaia}.

\begin{figure}
\includegraphics[width=1.0\columnwidth]{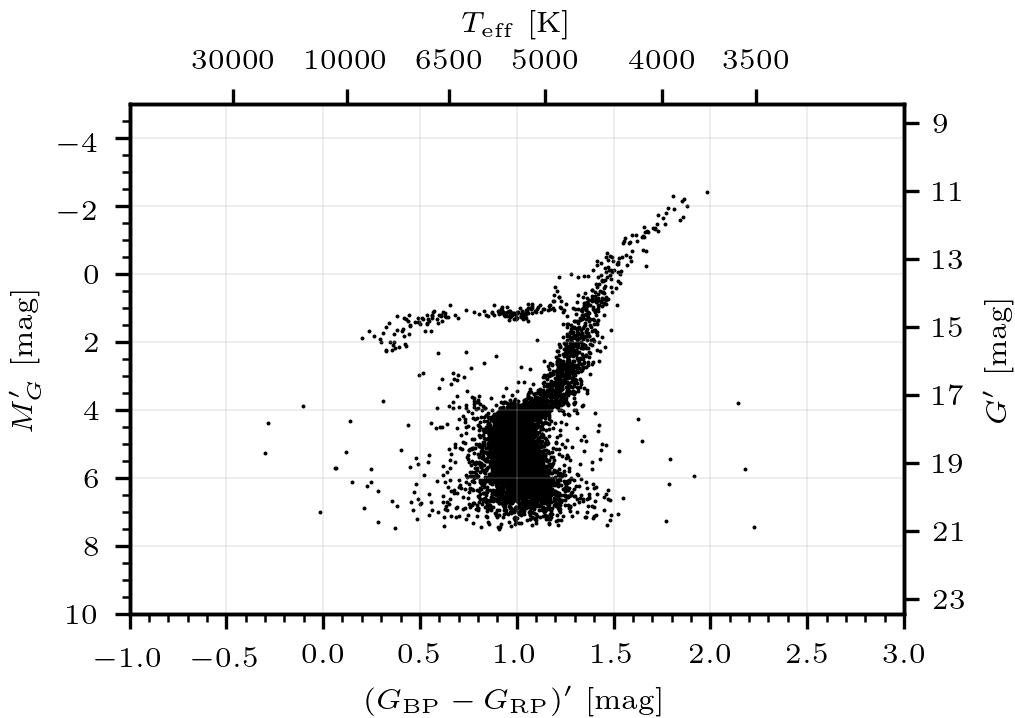}
\caption{Absolute magnitude without dust extinction correction, $M'_G$ versus observed $(\BPRP)'$ colour index for 7064 stars within 0.14 deg of the centre of NGC 3201, selected with the conditions of the query given in Appendix \ref{App0}. The absolute magnitude is computed assuming a heliocentric distance
$r_{\rm h} = 4.9\, {\rm kpc}$.}
\label{HR_NGC3201}
\end{figure}

We next take a simulation of the entire \textit{Gaia} catalogue, the 18th version
of the Gaia Object Generator \citep[GOG18;][]{2014AandA...566A.119L}.
This catalogue includes $\Approx 1.5$ billion sources with \Gband magnitude
$G\LessSim20$, so a pre-selection of a greatly reduced sample of tidal
stream candidates is necessary before we can computationally implement a
maximum-likelihood method to fit a tidal stream to the candidates.
We apply various pre-selection cuts as described in PM19 (Section 3.3):
(1) $G < 21$, to reduce faint stars with large errors; (2) parallax
$\pi>1/0.3$ mas, to eliminate foreground disc stars; a third cut in
PM19 that removed stars at low Galactic latitude is not applied here.

The fourth cut is the most important one, causing the greatest reduction in the number of star candidates. It is defined in Appendix C of PM19. Basically, we define a phase-space volume around the cluster orbit. By calculating the intersection of each star with this volume, we can choose stars close enough to the initial tidal stream model to be feasible stream stars in the final best-fitting model. We construct this volume by computing a bundle of orbits around the cluster orbit by variations of the current phase-space position of the cluster within observational errors, and variations of the Milky Way dark halo parameters. For the halo parameters, we randomly generate values following uniform distributions within the intervals: $\rho_{0 \rm dh} \sim (8 \pm 1) \pd{6}$ M$_{\odot}$ kpc$^{-3}$, $a_{1 \rm dh} \sim 20.2 \pm 4$ kpc, $a_{3 \rm dh} \sim 16.16 \pm 4$ kpc, and $\beta_{\rm dh} \sim 3.1 \pm 0.2$. The selection threshold and the above distributions are defined such that the pre-selection criterion is broad enough to remove very few true stream stars. We require the pre-selected stars to be in the interval going from a time 60 Myr in the past to 10 Myr in the future (shown as solid red line in Figures \ref{stream_NGC3201} and \ref{gog18_pre_short}) because the dense stellar foreground and large distance to the stream make it difficult to find stream stars outside this interval. Finally, the fifth cut removes stars within $1.5$ deg of the cluster centre, to remove stars that may still be bound to the cluster and are not part of the tidal stream.

 The GOG18 pre-selected stars after these cuts (a total of 486 664, as
listed in Table \ref{results_cuts}) are plotted in the left-hand panels of Figure
\ref{gog18_pre_short} as blue dots. The right-hand panels also shows as blue
dots the 1609 stars in our simulated tidal stream that pass the same
pre-selection cuts. Most of the other stars in our simulated stream
are eliminated because they are far from our orbital interval
from $-60$ to $10\, {\rm Myr}$, where detecting the
candidates is more difficult. We can see how within this interval, only a few simulated stream stars have not passed these cuts (grey dots) proving that our fourth cut does not bias the selection. The density of contaminating foreground stars is minimum in the range $\alpha\in\Range{80}{120}$ deg,
corresponding to the section of the stream closest to the Sun where the
proper motion is largest. Our pre-selection volume cuts out most of the
leading arm, as well as the distant ends of the simulated stream. These
cut out regions are far from the Sun, projected near the disc and with
proper motions that have a high density of foreground stars. We will
focus in the search for candidates in the portion of the tidal stream
defined by our cuts in this paper, although other stream stars are
expected to be found over the more extended, complete simulated stream
in future work.

\begin{table}
\caption[]{\small{Total number of stars in GOG18 and GDR2 and number
that pass each cut. For cuts 6 and 7, the number of stars left is shown
divided in six sky regions, where the tidal tail is seen under different
foreground conditions, and a different value of the threshold
$\chi_{\rm sel}$ defined in section \ref{subs:fsel} is used, specified
in units of
${\rm yr}^{3}\,{\rm deg}^{-2}\,{\rm pc}^{-1}\,{\rm mas}^{-3}$.
Numbers in parentheses for GOG18 indicate the expected number of stars
if GOG18 had the same number of pre-selected stars as GDR2 in each
region. Note that cut 7 is not used to obtain the best-fitting tidal stream
model, but only for the final selection of candidate stream members. The
number of selected stars combining all six regions is shown at the
bottom. }}
\begin{center}
\begin{tabular}{llrr}
\toprule
\multicolumn{2}{l}{\textbf{Pre-selection cut}}&\textbf{GOG18}&\textbf{GDR2}\\
\midrule
\multicolumn{2}{l}{All catalogue}&1510 398 719&1692 919 135\\
\multicolumn{2}{l}{(1)\,-\,(2)}&1490 962 149&1313 216 777\\
(4) &&492 983&250 764\\
(5) &&486 664&218 065\\
\midrule
\multicolumn{3}{l}{\textbf{Region (i) Disc foreground 1}}&\\
\midrule
(6) $\chi_{\rm sel} = 4.9\pd{-2}$&&2 (1)&18\\
(7) &&1 (0)&12\\
\midrule

\multicolumn{3}{l}{\textbf{Region (ii) Disc foreground 2}}&\\
\midrule
(6) $\chi_{\rm sel} = 7\pd{-3}$&&1 (0)&14\\
(7) &&1 (0)&8\\
\midrule

\multicolumn{3}{l}{\textbf{Region (iii) Stream}}&\\
\midrule
(6) $\chi_{\rm sel} = 5\pd{-3}$&&10 (4)&55\\
(7) &&8 (4)&51\\
\midrule

\multicolumn{3}{l}{\textbf{Region (iv) Dust}}&\\
\midrule
(6) $\chi_{\rm sel} = 5.38\pd{-4}$&&1 (0)&7\\
(7) &&0 (0)&6\\
\midrule

\multicolumn{3}{l}{\textbf{Region (v) Globular Cluster}}&\\
\midrule
(6) $\chi_{\rm sel} = 2.9\pd{-3}$&&11 (5)&75\\
(7) &&2 (1)&71\\
\midrule

\multicolumn{3}{l}{\textbf{Region (vi) Disc foreground 3}}&\\
\midrule
(6) $\chi_{\rm sel} = 6\pd{-3}$&&3 (1)&28\\
(7) &&2 (1)&22\\
\midrule

\multicolumn{3}{l}{\textbf{All regions combined}}&\\
\midrule
(6) &&28 (13)&197\\
(7) &&14 (6)&170\\
\bottomrule
\end{tabular}
\end{center}

\label{results_cuts}
\end{table}

\section{Selection and detection of the NGC 3201 stream stars from GDR2}\label{sec3}

  Our goal in this section is to obtain a best-fitting model of the stellar
stream associated with NGC 3201 varying parameters of the Milky Way
potential and the globular cluster kinematics, showing at the same time
that the kinematic data of the GDR2 catalogue proves the existence of
this stellar stream with a high degree of statistical confidence. A
list of candidate stellar stream members that are also photometrically
consistent with the NGC 3201 H-R diagram will be given. While each of
these candidates has some probability of being a false member (a
projected foreground or background star), our method relies on the
statistical detection and maximizes the stream-likelihood function based
on the number of candidates identified with a high membership
probability.

\subsection{Pre-selection of GDR2 stars}

 We first apply our pre-selection method to the GDR2 catalogue to reduce
the number of stars used to fit the stellar stream to a computationally
manageable level. This catalogue includes a total of $\Approx 1.7$
billion sources with parallaxes, sky coordinates and proper motions, and
$\Approx 7.2$ million sources with radial velocities. The number of
stars that pass each of our cuts defined in Section \ref{sec2} is
specified in Table \ref{results_cuts}, together with the same number for
the simulated catalogue GOG18. The first two cuts (1 and 2) eliminate a
small number of stars, and the main reduction is achieved in cut 4,
leaving $492\,983$  for GOG18 and $250\,764$ for GDR2. The difference of
a factor $\Approx 2$ between the two catalogues in the pre-selected
fraction is caused by imperfect modelling of the disc stellar population
or inaccurate estimation of observational errors in GOG18. Observational
errors are provided only for end-of-mission results in GOG18, while
GDR2 is based on data collected during the first 2 yr of the
\textit{Gaia} mission, so we have corrected the GOG18 errors as in
section 3.1 of PM19 but this correction may be inaccurate.
Incompleteness of the GDR2 catalogue in areas with lower than average
exposure or high stellar density and inaccurate modelling of dust
extinction in GOG18 may be other reasons for the difference of the
simulated and real catalogues. Cut 5 has again a relatively small
impact and removes more stars in the vicinity of NGC 3201 in GDR2 than
in GOG18 because the latter does not include globular clusters.

 Cuts 6 and 7 are applied only after the best fit to the stream
has been computed, to obtain a list of the most likely candidate stream
members. Cut 6 involves an accurate kinematic consistency with the best-fitting stream model, and cut 7 requires photometric compatibility with the
progenitor cluster H-R diagram, and will be discussed in detail in
section \ref{subs:fsel}.

\begin{figure*}
\includegraphics[scale=0.96]{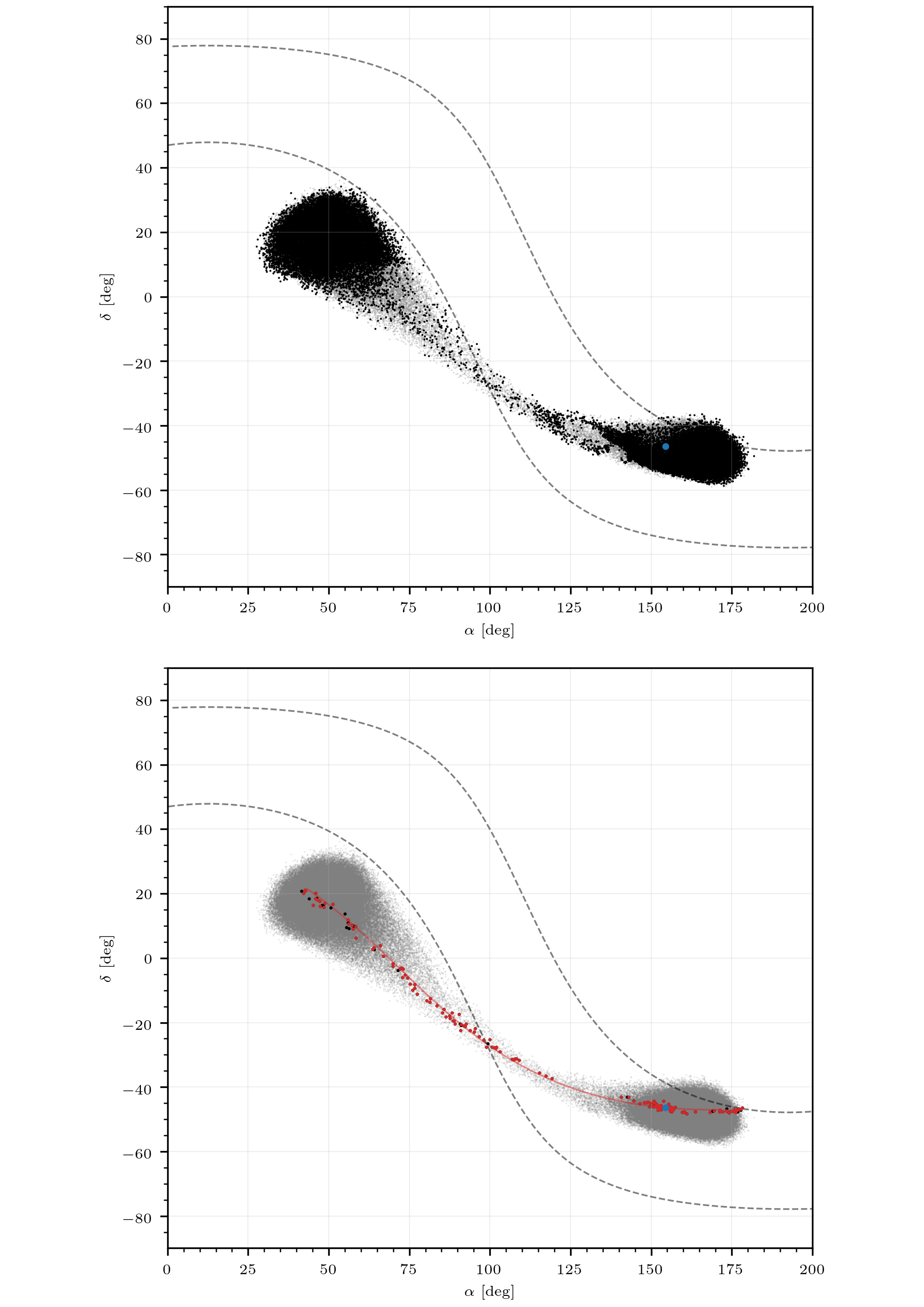}
\caption{\textit{Top}: Sky map in equatorial coordinates of the
pre-selected stars from GDR2 passing cuts 1, 2, 4, 5 (grey dots), which are the
stars that are judged to be roughly compatible with stream membership
before the fit is done, and are used to obtain the final fit. Stars that
are also compatible with the H-R diagram of NGC 3201 (passing cut 7) are
highlighted in black. An elongated overdensity in the range
$\alpha\Approx\Range{75}{100}$ deg and an obscured region by dust in
$\alpha\Approx\Range{120}{145}$ deg can be seen, which is much sharper
and clearer for stars passing cut 7, indicating the presence of the
stream. The blue dot is the present cluster position, and the grey lines
indicate Galactic latitude $b=\pm15$ deg.
\textit{Bottom}: Final selection of GDR2 stars. The grey small dots are the same
as in the top panel, black dots are stars compatible with the density
model of the best-fitting stellar stream (stars passing cut 6 with our
chosen values of $\chi_{\rm sel}$ in different regions), and red dots
are stars that are also compatible with the NGC 3201 H-R diagram (passing
cut 7). The red line is the best-fitting orbit of NGC 3201 from $-60$ to
$+10$ Myr from the present time.}
\label{ad_sel}
\end{figure*}

 The pre-selected stars, passing cuts 1 to 5, are shown in the top panel
of Figure \ref{ad_sel} as grey dots, in an equatorial coordinates sky
map with the Galactic latitude $b=\pm15$ deg shown as dashed lines,
and the position of NGC 3201 shown as a blue dot. These pre-selected
stars follow roughly the cluster orbit only because we have required
this in cut 4 when selecting stars consistent with a bundle of orbits
around that of NGC 3201, including uncertainties in the distance and
kinematic measurements and in the Galactic potential model. However,
the black dots in the top panel include the additional cut 7 (imposing
a consistent color with the NGC 3201 H-R diagram at the distance of the
stream model, see section \ref{subs:fsel}). The narrow band of these
black dots seen in the range $70 \LessSim \alpha \LessSim 100$ degrees is already a
visual evidence of the presence of the stream.

\subsection{Best fit to tidal stream from GDR2 kinematic data}

 We now apply the method of maximum likelihood to compute the best-fitting
parameters of the stream model, varying both orbital parameters of
NGC 3201 with the prior of the distance, radial velocity and proper
motion observational determinations, and parameters for the Galactic
halo determining the gravitational potential. The method we use is fully
described in PM19 and is based on an approximate calculation of a
likelihood function, computed from a stellar density of the tidal stream
inferred from our stream simulation, and from a model distribution
function of the foreground stars belonging to the general Milky Way
stellar populations. As explained in section 2.5.1 of PM19, the
simulation that is used to compute a model of the stream stellar
distribution is performed by following the trajectory of stars that are
initially in orbits with a significant escaping probability (obeying
equation 21 in PM19), derived from a tidal radius
$r_t = R_{\rm c} [M_{\rm gc}/(3M_t)]^{1/3}$, where
$R_{\rm c}$ is a characteristic orbital radius of the cluster,
$M_t=5.2\pd{11} M_{\Sun}$ is the total Galaxy mass (this was written as $M$ in equations
19 and 20 of PM19), and $M_{\rm gc}$ is the globular cluster
mass given in Table \ref{NGC3201_DH}. The true tidal radius is somewhat
smaller than $r_t$ because only the Galaxy mass interior to $R_{\rm c}$
counts for generating the tidal stress on the cluster, but in practice
we adjust $R_{\rm c}$ so that fewer than 30 per cent of the stars that escape
are missed because of not including them in our fast simulations that
follow only stars with a high escape probability.
We have chosen on this basis $R_{\rm c}= 15$ kpc for NGC 3201 in this
paper.

 We do not apply this method using the entire \textit{Gaia} stellar
catalogue, which would be prohibitively expensive computationally, but
we use only the pre-selected stars to compute our likelihood function.
This essentially neglects the possibility that any stars outside our
pre-selected sample might be stream members. The free parameters we use
and their best-fitting results are listed in Table \ref{results_param}:
the fraction of stars $\tau$ in the stellar stream, parameters of the
halo density profile ($\rho_{0 \rm dh}$, $a_{1 \rm dh}$, $a_{2 \rm dh}$,
and $\beta_{\rm dh}$), and the heliocentric distance, radial velocity
and proper motions of the globular cluster. Gaussian priors from
the observational results listed in Table \ref{NGC3201_DH} are used for
the cluster present phase-space coordinates, while the remaining
parameters are given uniform priors wide enough to be unimportant for
the results. Errors listed in Table \ref{results_param} are from the
diagonal elements of a full covariance matrix of all the free
parameters, computed from the second derivatives of the posterior
function. A few other derived parameters for the NGC 3201 orbit, and
statistical measures defined in PM19, are also included in Table
\ref{results_param}.

\begin{table}
\caption[]{\small{Best-fit parameters obtained for the NGC 3201 orbit and
the Galactic dark halo, using the GDR2 pre-selected data.}}
\begin{center}
\begin{tabular}{lllc}
\toprule
\multicolumn{2}{l}{\textbf{Statistical parameters}}\\
\midrule
$\varLambda$&&$364.61$\\
$\tau$&&$(2.74\pm0.35)\pd{-4}$\\
$Q$&&$17.95$\\
\midrule
\multicolumn{3}{l}{\textbf{Best-fitting kinematics of NGC 3201}}\\
\midrule
$r_{\rm h}$&\units{kpc}&$4.79\pm0.05$\\
$v_r$&\units{km s$^{-1}$}&$494.304\pm0.14$\\
$\mu_\delta$&\units{mas yr$^{-1}$}&$-1.9859\pm0.0014$\\
$\mu_\alpha$&\units{mas yr$^{-1}$}&$12.1048\pm0.0022$\\
\midrule
\multicolumn{3}{l}{\textbf{Dark halo best-fitting parameters}}\\
\midrule
$\rho_{0 \rm dh}$&\units{M$_{\odot}$ kpc$^{-3}$}&$(6.87\pm0.09)\pd{6}$\\
$a_{1 \rm dh}$&\units{kpc}&$18.67\pm0.27$\\
$a_{3 \rm dh}$&\units{kpc}&$16.29\pm0.25$\\
$\beta_{\rm dh}$&-&$2.845\pm0.035$\\
\midrule
$q$&-&$0.87\pm0.02$\\
$q_{\varPhi}$&-&$0.94\pm0.01$\\
$c_{200}$&-&$9.14\pm0.2$\\
$r_{200}$&\units{kpc}&$202.2\pm5.2$\\
$M_{200}$&\units{M$_{\odot}$}&$(9.71\pm0.75)\pd{11}$\\
\midrule
\multicolumn{3}{l}{\textbf{Derived NGC 3201 orbital parameters}}&\\
\midrule
$r_{\rm peri}$&\units{kpc}&$7.67\pm0.03$&\\
$r_{\rm apo}$&\units{kpc}&$37.62\pm1.41$&\\
$L_{z}$&\units{km s$^{-1}$ kpc}&$2728.8\pm18.4$&\\
\bottomrule
\end{tabular}
\end{center}
\label{results_param}
\end{table}

 Our results can be described according to the following three points:

\begin{enumerate}

 \item A tidal stream of NGC 3201 is detected at a very high confidence
level. This is inferred by maximizing the likelihood function, which
essentially corresponds to maximizing the overlap of the stream
phase-space distribution model with the stellar distribution in our
pre-selected data. The value of $\tau\sim 3\pd{-4}$ we find for
our best fit, which is the fraction of stars in our pre-selected sample
that belong to the stream if the best-fitting model is correct, has a
relatively error of only 12 per cent, so it is greater than zero with a very
high statistical significance. Note that this number does not have a
useful physical interpretation because it depends on our pre-selection
method, and also on the complex details of the selection of the \textit{Gaia}
catalogue. In addition, the value of the statistic $\Lambda$ indicates
the confidence level at which the presence of the stream is detected,
as explained in PM19 (Section 2.1). When $\Lambda>6.6$, the existence
of the stream is confirmed at the 99 per cent confidence level compared to the
null hypothesis that no stream is present. The large value of $\Lambda$
implies a very high detection statistical significance.

 \item The best-fitting orbit of NGC 3201 matching the detected stream is
remarkably close to the orbit that is derived exclusively from the
independent observational determinations of the cluster phase-space
coordinates. The statistical parameter $Q$ quantifies the deviation of the best-fitting
present phase-space coordinates of NGC 3201 from the observational
determinations in Table \ref{NGC3201_DH}. Its expected value is the number of parameters of the globular cluster orbit that are fitted (in this case 4, as given in Table \ref{results_param}). The larger obtained value $Q\simeq 18$ is mostly due to the deviation of the best-fitting from the
observed proper motion along right ascension, a $4.3\sigma$ deviation.
We note that this deviation, while significant compared to the small
statistical measurement errors of the GDR2 proper motion of the
globular cluster, are actually less than 0.2 per cent of the proper motion.
This small deviation may be caused by underestimates of errors provided
by \citet{2018AA...616A..12G} for globular cluster proper motions,
obtained by averaging measuremens of a large number of member stars.
Both \citet{2019MNRAS.484.2832V} and \citet{2019MNRAS.482.5138B} provide
bigger uncertainties compatible with our results.

 \item In our Galactic potential model where only halo parameters are
allowed to vary, our best-fit result for these parameters (listed in
Table \ref{results_param}) has relatively small errors, and is
remarkably close to our best-fitting M68 stream model from PM19 (with values
listed in Table \ref{NGC3201_DH}). In particular, the axis ratio
of the halo mass density distribution $q=0.87$ has an error of only 2 per cent,
and is in very good agreement from the two streams. However, these small
errors are of course the result of assuming a fixed model for the disc
and bulge parameters. If these are allowed to vary, parameter
degeneracies arise which are expected to increase the error in the halo
parameters by a large factor. In addition, the computation of the uncertainties by the second derivative of the likelihood function underestimates the errors of the free parameters when the posterior function is not smooth over the entire range of possible parameter values. Mainly, this occurs for the halo parameters.

\end{enumerate}

\subsection{Visualization of stream including photometric selection}

  Apart from obtaining the best fit to the stream using the kinematic
data and noting the large value of $\Lambda$ in Table
\ref{results_param}, there is an alternative way to test the reality of
the stream: we can search for stars with photometry that is compatible
with the H-R diagram of NGC 3201, assuming that they are stream members
and are at the distance predicted by the stream model. These stars
should be distributed along a narrow region of phase-space corresponding
to the stream when compared to the whole pre-selected sample.

 We define a seventh cut (7) that selects stars consistent with the
H-R diagram of the globular cluster NGC 3201, obtained directly from the
GDR2 catalogue as described in Appendix \ref{App0}. The method is the
same as that described in PM19 (section 3.5 and Appendix D), which
basically defines a density model in the H-R diagram based on the
cluster member stars, and then selects stream stars with a position in
the H-R diagram above a threshold density (we use in this paper a
threshold $P_{\SM{CR}}\GtrSim0.035$ mag$^{-2}$, as defined in equation
D5 of PM19). An important difference from PM19 is, however, introduced:
we take into account dust extinction, which in this case is important
because the cluster is located close to the Galactic plane in a region
of moderately high extinction, and the stream is also affected by
varying amounts of extinction over its long extent. The corrections
applied to stars to both the magnitude and color for dust
extinction and reddening are from \cite{2011ApJ...737..103S}, known as
the SF dust extinction map model, and is described in detail in
Appendix \ref{App1}. Note that this dust correction is obtained from the
SF model assuming that all the dust is foreground to the stars, an
assumption that is valid in most cases for stars with low \textit{Gaia} parallax
except when looking at very low Galactic latitude (in which case
extinction is very high anyway).

\begin{figure}
\begin{tabular}{c}
\includegraphics[width=0.95\columnwidth]{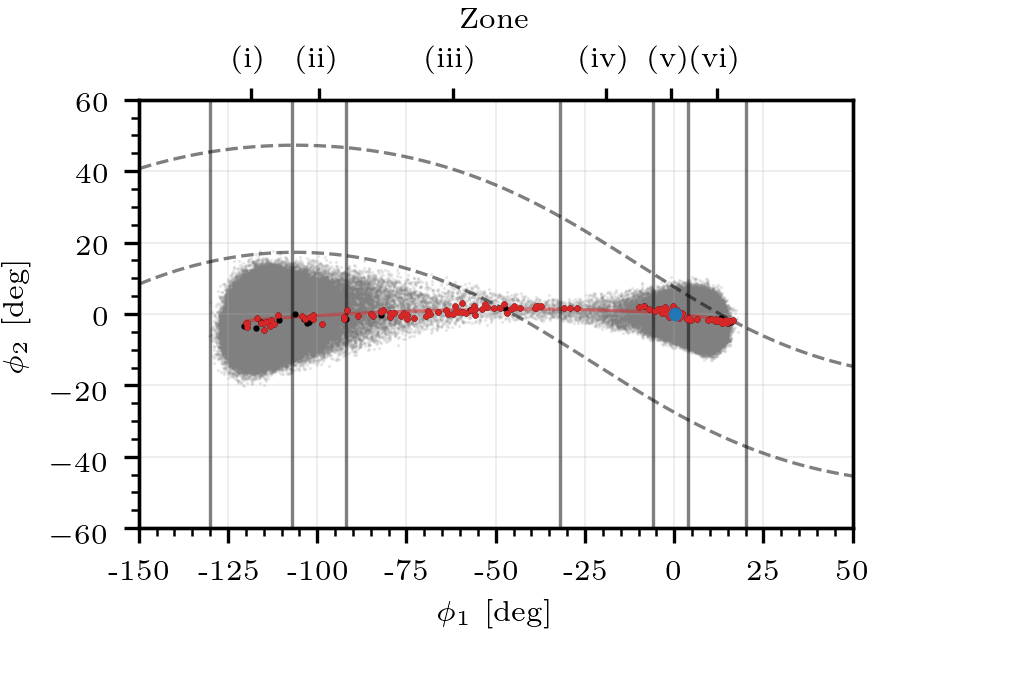}\\
\vspace{-0.75cm}\\
\includegraphics[width=0.95\columnwidth]{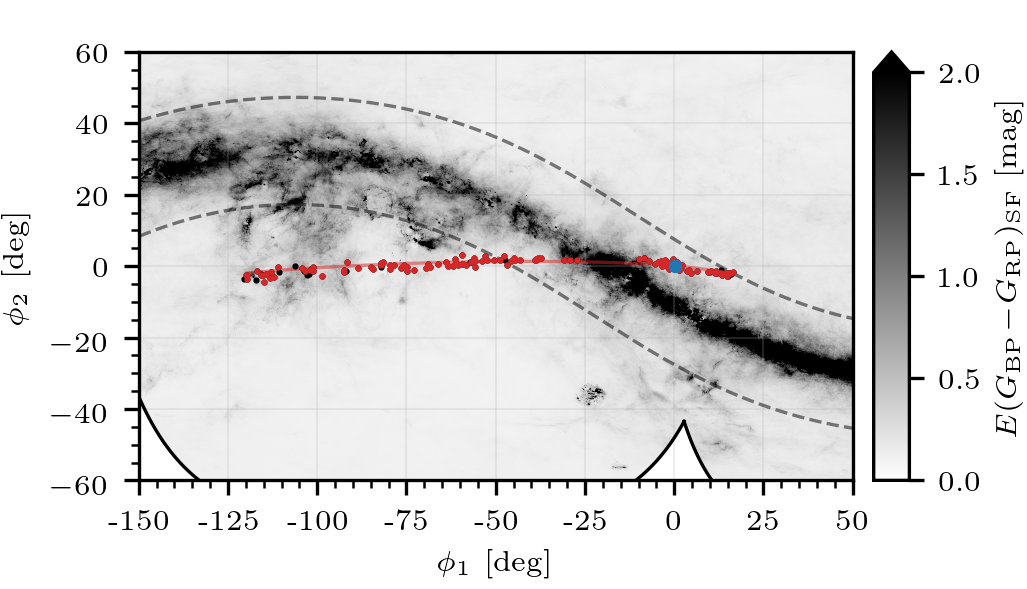}
\end{tabular}
\caption{Same as bottom panel of Figure \ref{ad_sel}, expressed in a 
stream coordinate system with the $\phi_1$ axis following the cluster
orbit, as specified in Appendix D. \textit{Top:} The solid vertical lines
mark selection zone limits. \textit{Bottom:} Colour excess of the
$\BPRP$ colour index from the SF extinction map
\citep{2011ApJ...737..103S}, a recalibrated version of the SFD
extinction map \citep{1998ApJ...500..525S}. Only the best-fitting stream
candidate member stars after cut (6) are shown here.}
\label{dust}
\end{figure}

 Among all the pre-selected stars shown as grey dots in
Figure \ref{ad_sel}, those that are in addition compatible with the
NGC 3201 H-R diagram (cut 7) are shown as black dots of larger size in
the top panel. The large blue dot is the present position of NGC 3201.
An elongated overdensity which is narrower than the whole pre-selected
sample is clear in the range $70 \LessSim \alpha \LessSim 100$ deg. We note that
the region very close to the Galactic plane has very few stars that
pass this cut 7. The reason is the very large extinction present in
this region.

 The stream and selected stars are better visualized by plotting these
maps in rotated spherical coordinates, where the angle $\phi_1$ varies
along a major circle that approximately follows the stellar stream, and
the angle $\phi_2$ is a polar angle from the axis perpendicular to this
major circle. The bottom panel of Figure 4 shows stars that pass not
only cut 7, but also cut 6 which requires kinematic consistency with
the best-fitting stream model (described in detail in the next subsection).
This is replotted in the stream coordinates $(\phi_1,\phi_2)$ in 
Figure \ref{dust} in the top panel. The bottom panel of Figure
\ref{dust} adds the dust extinction map of the SF model
\citep{2011ApJ...737..103S}. We can see that the region of the stream
with an absence of stars compatible with all our cuts coincides with
the region of highest dust extinction. The globular cluster NGC 3201
is seen at a moderately low, northern Galactic latitude, where dust
extinction is close to 1 magnitude, and the trailing arm is the one
that passes closest to the Solar System and is therefore most visible
to us. This trailing arm crosses the Galactic disc and reappears on the
southern Galactic hemisphere, where most of the stream candidates in
the GDR2 catalogue can be identified.

\subsection{Final stream star selection including photometry}
\label{subs:fsel}

 The final procedure in our study of the tidal stream is to select a
list of stars that are most likely to be stream members, using both the
stellar density of the stream model that gives the maximum likelihood,
and the photometric condition of consistency with the cluster H-R
diagram. First, cut 6 selects the stars with kinematic variables that
are compatible with the best-fitting stream model, and then cut 7 restricts
our final list to stars compatible with the H-R diagram.

 The stream phase-space density model is constructed as in PM19,
similar to the way we evaluate the likelihood function, from the
superposition of several Gaussian distributions along the stream. We
compute the phase-space density at the phase-space position of each of
the pre-selected stars for the best-fitting density model, convolving it
with the observational errors of the phase-space coordinates, and we
select stars with a value of this convolved phase-space density above a
threshold $\chi_{\rm sel}$, which is expressed in units of
${\rm yr}^{3}\,{\rm deg}^{-2}\,{\rm pc}^{-1}\,{\rm mas}^{-3}$. The
procedure is also done with the simulated stars in the GOG18 catalogue,
and the threshold is chosen in each zone so that a small number of GOG18
stars are selected as stream members (this small number obviously
represents our noise level of false candidates because there are no
streams in the GOG18 simulation).

 In practice, the stream we are analysing is very long and the different
regions of the sky over which it is projected have very different levels
of foreground contamination. To optimize our stream candidate list, we
divide the sky into six different zones and use a different value of
$\chi_{\rm sel}$ in each one. We set the zone limits in the stream
coordinate $\phi_1$, defined to be the angle along a major circle that
is approximately followed by the stream. The transformation from
equatorial to these stream coordinates is given in Appendix \ref{App3}.
The six regions, shown in the top panel of Figure \ref{dust}, are
defined as follows:
\begin{center}
\begin{tabular}{llcl}
(i) & \textit{Disc foreground 1}: & $\!\!-130 \leqslant \phi_1 < -107 $ &
$\!\units{deg}$\\[1em]
(ii) & \textit{Disc foreground 2}: & $\!\!-107 \leqslant \phi_1 < -92$ &
$\!\units{deg}$\\[1em]
(iii) & \textit{Clean stream}: & $\!\!-92 \leqslant \phi_1 < -32$ &
$\!\units{deg}$\\[1em]
(iv) & \textit{High dust}: & $\!\!-32 \leqslant \phi_1 < -6$ &
$\!\units{deg}$\\[1em]
(v) & \textit{Globular cluster}: & $\!\!-6 \leqslant \phi_1 < 4$ &
$\!\units{deg}$\\[1em]
(vi) & \textit{Disc foreground 3}: & $\!\!\quad 4 \leqslant \phi_1 \leqslant 20$
& $\!\units{deg}$\\[1em]
\end{tabular}
\end{center}
We list in Table \ref{results_cuts} the number of stars of the GOG18 and
GDR2 catalogues that pass cuts 6 and 7, and the value of the selection
threshold $\chi_{\rm sel}$ we choose for each zone.

 In zones (i), (ii), and (vi), the density of foreground stars is very
high because the proper motions and parallaxes of most disc stars are
small and cannot be distinguished from the stream stars. This makes our
cuts less effective at reducing the number of stars in our pre-selected
sample. Taking these three zones together, we select 60 stars from GDR2
that pass cut 6, while only six are found by chance in GOG18 with the same
values of $\chi_{\rm sel}$. This suggests most of the 60 stars found in
this zone are real stream members, even before applying our cut 7.
Actually, the number of six stars found in GOG18 in these three zones is
an overestimate of the number of false candidates we should expect in
GDR2, because the number of stars that are pre-selected in GOG18 is
larger than in GDR2 (as seen in Table \ref{results_cuts} in the total
number of pre-selected stars after cut 5). A more reasonable estimate
of the expected number of false candidates in GDR2 is obtained by
correcting the number found in GOG18 according to the ratio of pre-selected
stars after cut 5 in GOG18 and GDR2 in each of our six zones. This
corrected estimate is written in parenthesis after the GOG18 number in
each zone. For these three zones, the expected number of false candidates
is reduced to 2 or 3. When applying in addition cut 7, the stream
candidates are reduced to 42. In GOG18, the noise candidates are not
reduced very much by cut 7 because most of the contaminating disc stars
in these regions have colours that happen to be compatible with the
NGC 3201 H-R diagram when the model stream distance is assumed.

 Zone (iv) is highly obscured by dust (see Figure \ref{dust}). Only seven stars are selected, of which six are compatible with the cluster H-R
diagram. These stars are actually all located at the edges of zone
(iv), where the dust extinction is not so high, and we therefore think
they are most likely true stream members.

 Zone (v) corresponds to the vicinity of NGC 3201. We select 75 stars,
of which 71 are compatible with the cluster H-R diagram. In contrast, in
GOG18 only 11 stars pass cut 6, of which only 2 pass cut 7, indicating
that most of our final 71 stars from this zone are truely associated
with the cluster. Our cut 5 removes stars only within an angle of 1.5
deg from the centre of NGC 3201, and there may still be some cluster
member stars outside this angle that are bound to the cluster; in fact,
the 71 stars in zone (v) we include in our final selection are rather
concentrated towards the cluster. It is in general ambiguous to separate
stars that are still bound from those that are already considered as
stream members.

\begin{figure}
\includegraphics[width=1.0\columnwidth]{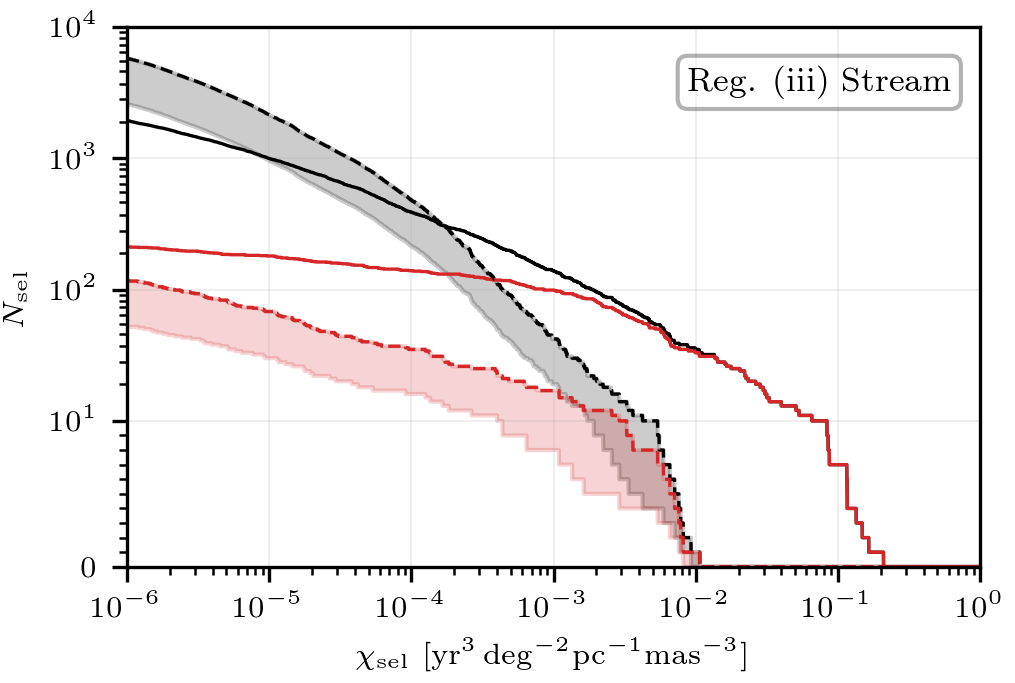}
\caption{Number of selected stars $N_{\rm sel}$ in zone (iii) as a
function of the selection threshold $\chi_{sel}$.
 {\it Black solid line:} number of GDR2 stars compatible with the
best-fitting stream model, passing cut 6.
 {\it Red solid line} number of stars also compatible with NGC 3201 H-R
diagram (cut 7). {\it Dashed lines:} same quantities for GOG18 stars.
Shaded areas mark the reduction due to correcting for the larger number
of pre-selected stars in GOG18 than in GDR2.}
\label{sel_res}
\end{figure}

 Finally, zone (iii) is the cleanest because it contains the part of the
stream that is closest to us, located at $\Approx3.2$ kpc from the Sun,
with proper motions that are larger than those of most foreground stars.
Dust extinction is also relatively low. To analyze the selection in this
particularly favourable zone more carefully, Figure \ref{sel_res} shows
the number of selected stars, $N_{\rm sel}$, as a function of the
selection threshold for this zone. The solid black line is the number of
GDR2 stars selected for each threshold value, and the red one shows the
number that are also compatible with the NGC 3201 H-R diagram. The two
lines coincide up to $N_{\rm sel}\Approx100$, whereas at larger numbers
(or lower $\chi_{\rm sel}$) we start having many stars passing cut 6
which do not pass cut 7. This implies that for $N_{\rm sel} \lesssim
100$, most selected stars should be true members. Dashed lines are
the same for GOG18 stars. The shaded area below the dashed stars
indicates the range by which the number of stars found in GOG18 drops
because of the correction applied for the fact that GOG18 contains more
stars than are detected in GDR2 in our pre-selected sample. As
$\chi_{\rm sel}$ is dropped, the number of selected stars in GOG18 rises
to an increasing fraction of the ones found in GDR2. For our chosen value of $\chi_{\rm sel}$ in zone (iii), we select 55 stars of which 51 also pass cut 7, expecting a number of wrongly
selected stars of only $\sim$ 4.

 Our final selection over all regions contains 170 stars compatible with
our best-fitting stream model and the cluster H-R diagram. Our GOG18
estimate of the foreground contamination predicts that the number of
false members in this list is probably as low as $\sim 6$. These stars
are plotted as red dots in the bottom panel of Figure \ref{ad_sel} and
in Figure \ref{dust}, with black dots being for stars that pass only
cut 6 but not cut 7. We also show the best-fitting orbit of NGC 3201 as a
red solid line.

\begin{figure}
\begin{tabular}{c}
\includegraphics[width=0.95\columnwidth]{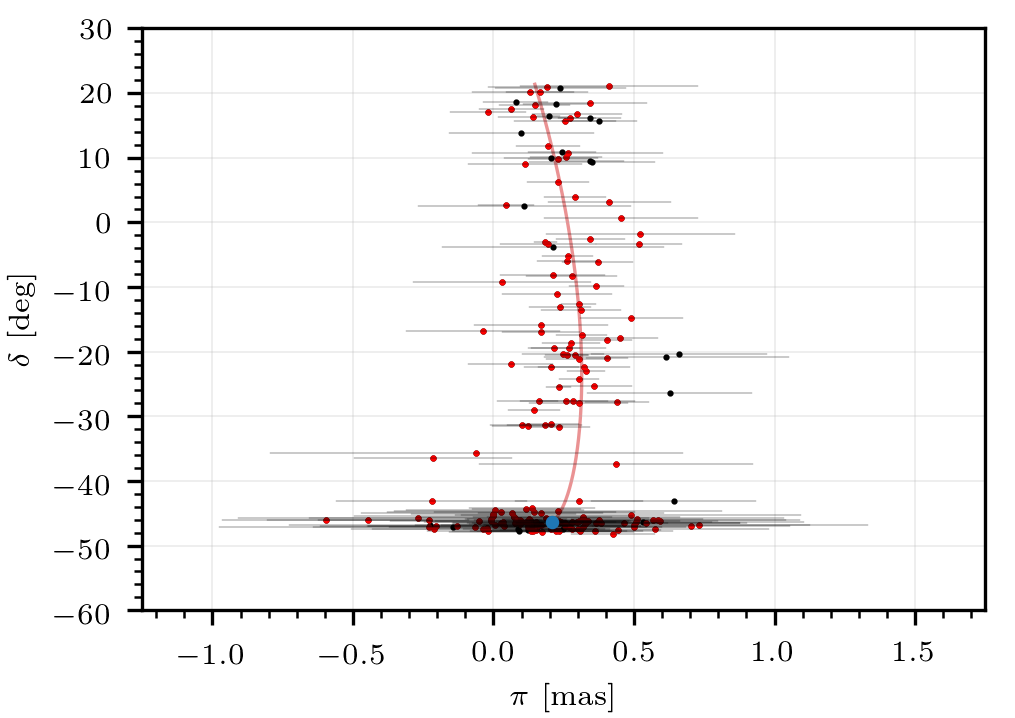}\\
\includegraphics[width=0.95\columnwidth]{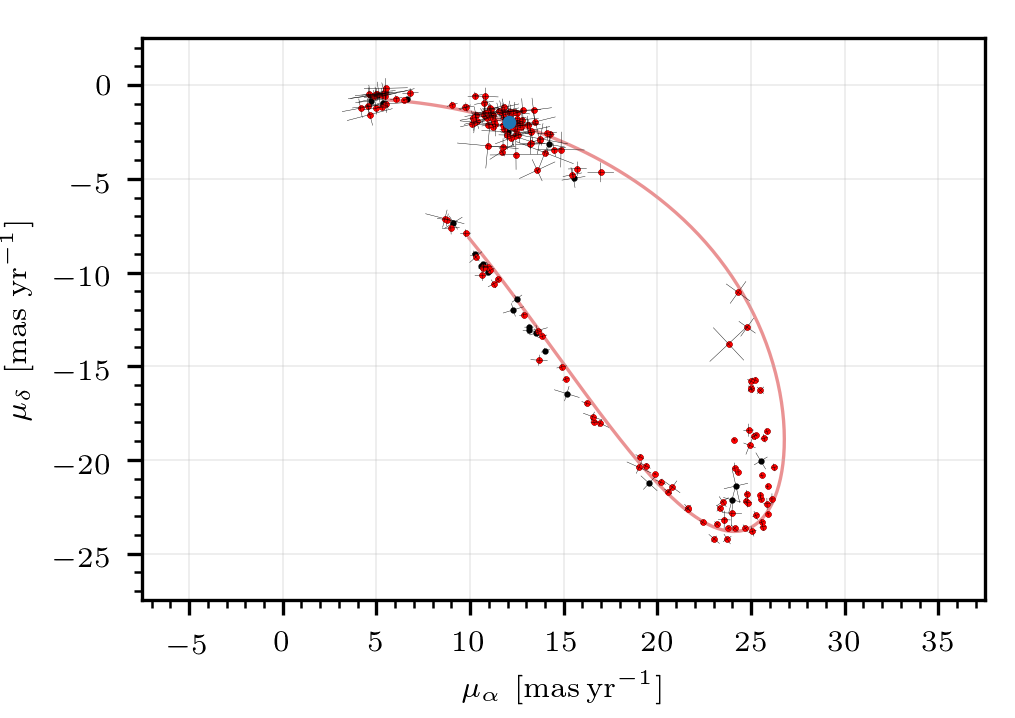}\\[0.5em]
\includegraphics[width=0.95\columnwidth]{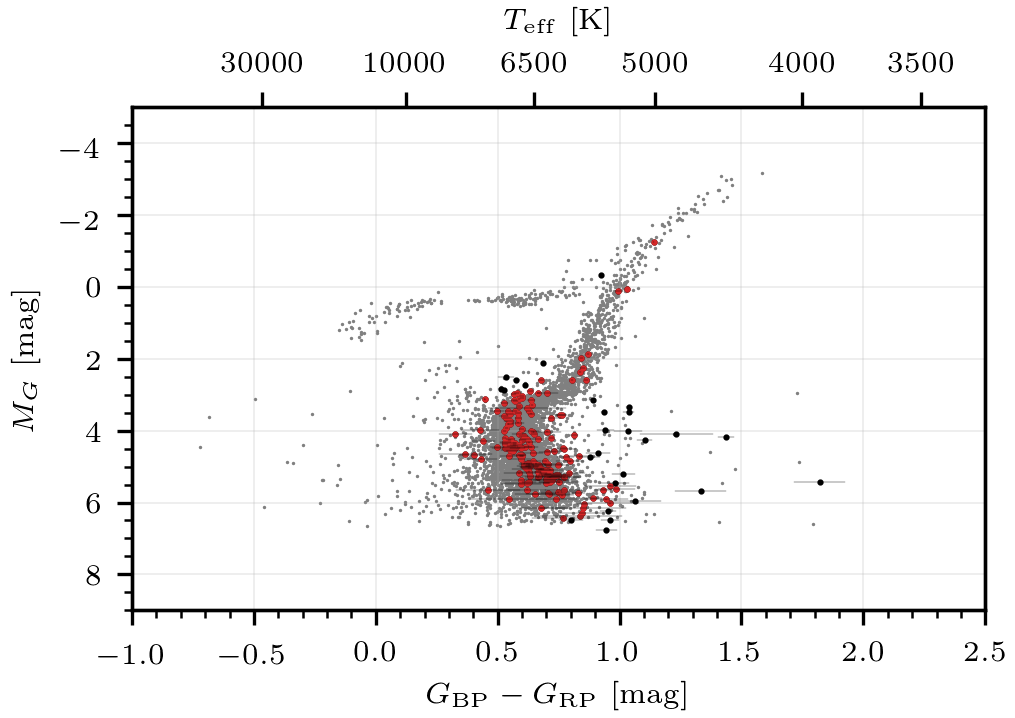}\\
\end{tabular}
\caption{Parallax versus declination (top), proper motions
(middle), and dust-corrected $\BPRP$ colour versus absolute
$\Gband$ magnitude (bottom) for the 197 stars in our final
selection after cut 6, with 170 of them passing also cut 7 shown in red,
and the remaining 27 in black. Thin black lines are observational
errors. Small grey dots in bottom panel are stars in NGC 3201.}
\label{ps_sel}
\end{figure}

 Figure \ref{ps_sel} shows other variables for these same stars:
parallax versus declination in the top panel, proper motions in the
middle panel, and the H-R diagram in the bottom panel (with cluster
members as small grey dots). Observational errors are indicated as thin
black lines. The top panel shows that the parallax is not a very useful
discriminant because the distance to the stream is too large for present
\textit{Gaia} uncertainties, but is nevertheless of some use and fits well the
expected orbit. Proper motions are the most valuable information when
detecting and modelling the stream. We note that, remarkably, using only
the kinematic selection for stream members (up to cut 6, which are both
the red and black dots), and inferring their absolute magnitude from
the stream model distance and correcting for dust, we reproduce the
H-R diagram of the cluster notably well.

 The list of our final 170 stream member candidates is in Appendix
\ref{App2}, with their measured coordinates, parallax and proper motion,
colour index $\BPRP$, and $\Gband$ magnitude. Only one star has a \textit{Gaia}
radial velocity in GDR2 of 499.29 km s$^{-1}$, which we find to be in agreement with the
orbital radial velocity prediction of 498.75 km s$^{-1}$.
We have checked the RAVE DR5 \citep{2017AJ....153...75K} and the
LAMOST DR4 \citep{2015RAA....15.1095L} catalogues, and have not found
any matches to this list.

\section{Conclusions}\label{sec4}

The method presented in PM19 is applied to search for a tidal stream associated with the globular cluster NGC 3201. This method identifies the stellar stream by statistically detecting star overdensities in a sample of observational data with respect to a phase-space density model of the Milky Way. For the best-fitting location of the globular cluster and the parameters of the gravitational potential, we construct a density model of the stream and select the stars with the highest intersection. Finally, we present as a final selection the stars that are also compatible with the H-R diagram of the progenitor cluster.

We detect a total of 170 stars candidates along the leading and the trailing arm of the stellar stream, extending over $\Approx$140 deg on the sky, from 40 to 180 deg in the Southern Galactic hemisphere, following an orbit of $L_z = 2728.8 \pm 18.4$ km s$^{-1}$ kpc. The clearest section of the stream spans from 70 to 105 deg, close to the Galactic disc, at about 3.2 kpc from the Sun. This section coincides with the known stellar stream Gj\"oll discovered by \citet{2019ApJ...872..152I} using the method Streamfinder \citep{2018MNRAS.477.4063M,2018MNRAS.478.3862M}, which spans from 70 to 90 deg in the Southern Galactic hemisphere, at $3.38\pm 0.1$ kpc from the Sun, following an orbit of $L_z = 2721 \pm 159$ km s$^{-1}$ kpc. This association based on \textit{Gaia} phase-space, colours, and magnitudes together with the chemical tagging of stars in the stream to NGC 3201 \citep{2020ApJ...901...23H}, proves that Gj\"oll is a section of the trailing tail of NGC 3201.

 Our best-fitting parameters are consistent with the observations of NGC
3201 and provide a consistent model of the Milky Way. Even so, our
computation underestimates the uncertainties of the halo parameters and
cannot be considered representative of our current understanding of the
density and potential of the Galaxy. In a future study, we will be using
a combination of several streams to study the constraints that can be
set on the Milky Way potential, especially the shape of the dark matter
halo, by fitting models with sufficient parametric freedom on all the
Galactic components to the observed data on all the stellar streams. Our
success in detecting this stream opens the possibility to detect many
more fainter stellar streams associated with globular clusters, as the
Gaia data improve and the separation from foreground stars becomes more
efficient.


\section*{Acknowledgements}

We would like to thank Holger Baumgardt for useful comments about the mass of NGC 3201.
This work has been supported by Spanish grants MDM-2014-0369 and
CEX2019-000918-M, for the Unit of Excellence Maria de Maeztu award to the
ICCUB. Use of data from the European Space Agency (ESA) mission
\textit{Gaia} (\url{https://www.cosmos.esa.int/gaia}) was made, processed
by the \textit{Gaia} Data Processing and Analysis Consortium
(DPAC, \url{https://www.cosmos.esa.int/web/gaia/dpac/consortium}). Funding
for the DPAC has been provided by national institutions, in particular, the
institutions participating in the \textit{Gaia} Multilateral Agreement.


\section*{Data Availability}

The data underlying this article are available in the article and in its online supplementary material.



\bibliographystyle{mnras}
\bibliography{bib/ref.bib}



\clearpage
\newpage 

\appendix

\newcounter{defC}
\setcounter{defC}{1}
\newcounter{prpC}
\setcounter{prpC}{1}


\section{Colour-Magnitude diagram of NGC 3201 from GDR2}\label{App0}

\noindent We reproduce here the ADQL query we have used to obtain the
photometry of all GDR2 stars in the $G$, $G_{\rm BP}$, and
$G_{\rm RP}$ passbands in a circle of radius $0.14$ deg centred on
NGC 3201, which yields 7064 stars:

\begin{lstlisting}[language=SQL, showspaces=false, basicstyle=\ttfamily, numbers=left, numberstyle=\tiny, commentstyle=\color{gray}, breaklines=true, keywords={AS, SELECT, FROM, WHERE, BETWEEN, IS, NOT, NULL, AND, CONTAINS, POINT, CIRCLE, COUNT}, keywordstyle=\color{mymauve}, stringstyle=\color{blue}, commentstyle=\color{mygreen}, xleftmargin=4.6mm]
SELECT bp_rp, phot_g_mean_mag, phot_bp_mean_flux, phot_bp_mean_flux_error, phot_rp_mean_flux, phot_rp_mean_flux_error, phot_g_mean_flux, phot_g_mean_flux_error 
FROM gdr2.gaia_source 
WHERE 1 = CONTAINS( POINT('ICRS', ra, dec), CIRCLE('ICRS', 154.3987, -46.4125, 0.14) ) 
AND parallax BETWEEN -1.6 AND 1.4 
AND SQRT((pmra-8.3344)*(pmra-8.3344) + (pmdec+1.9895)*(pmdec+1.9895)) <= 0.7 
AND bp_rp IS NOT NULL;
\end{lstlisting}

\begin{flushleft}
\phantom{a}
Host server: \url{https://gaia.aip.de/}\\
Description of the \texttt{gaia\_source} table: \url{https://gea.esac.esa.int/archive/documentation/GDR2/Gaia_archive/chap_datamodel/sec_dm_main_tables/ssec_dm_gaia_source.html}\\
\end{flushleft}


\section{Dust extinction correction}\label{App1}

To select stars that are consistent with the H-R diagram of NGC 3201, the
$\BPRP$ colour index and the $\Gband$ magnitude observed by \textit{Gaia} need to be corrected for the effects of dust extinction, both for the
cluster stars and the candidate stream stars. In general, for any observed
colour index $M'$, the corrected colour index $M$ is computed by subtracting
the colour excess $E_M$,
\begin{equation}
M = M'-E_M ~.
\end{equation}
We use the colour excess $E_{_{\BV}}$ for $\BV$ colour predicted by the
Galactic dust model of \citet{2011ApJ...737..103S}, known as the SF
model. This is the same as the colour excess model of
$\BV$ from \citet{1998ApJ...500..525S} reduced by a factor 0.86. The $\BV$ colour of stars can be related to the \textit{Gaia} colour
$\BPRP$, for most common stellar metallicities and gravities, using the
approximate expression of \citet{2010AandA...523A..48J}, from their
Table 3:
\begin{multline}
\BPRP = 0.0981 + 1.429\,(\BV) \\- 0.0269\,(\BV)^2 + 0.0061\,(\BV)^3 ~.
\end{multline}
We also follow the approximation of \citet{2010AandA...523A..48J} that
the dust extinction colour excess runs nearly parallel to this
colour-colour relation. Neglecting the small coefficients of the second- and third-order terms in $\BV$, we can use the simple approximation
\begin{equation}
E_{\BPRP} = 1.429\,E{_{\BV}} ~.
\end{equation}
The extinction correction in the $\Gband$ magnitude $A_{\SM{G}}$ can be
approximately expressed in terms of the colour excess $\BPRP$. We use
the expression calibrated at a typical dust extinction
$A_{\lambda = 550\, {\rm nm}} = 1$ mag, given in table 13 of
\citet{2010AandA...523A..48J}:
\begin{equation}
A_{\SM{G}} = 1.98 E\var{\BPRP} ~.
\end{equation}


\section{Final Candidate Stream Member Stars}\label{App2}

The selected stars from GDR2 catalogue after all our cuts from 1 to 7
are applied, which are our final list of best candidate stream members,
are listed in Table \ref{sel0}. Only one star in this list, star number
144, has a radial velocity:

\begin{center}
\begin{tabular}{rrrr}
\toprule
\multicolumn{1}{c}{N}&\multicolumn{1}{c}{source\_id}&\multicolumn{1}{c}{$v_r$}&\multicolumn{1}{c}{$\epsilon_{v_r}$}\\
&&\multicolumn{1}{c}{\units{km s$^{-1}$}}&\multicolumn{1}{c}{\units{km s$^{-1}$}}\\
\midrule
144&5365576065920333440&499.29&1.29\\
\bottomrule
\end{tabular}
\end{center}

\begin{table*}
\caption[]{\small{List of candidate stream member stars from the GDR2
catalogue, having passed all our 7 cuts. They are compatible with the
best-fitting phase-space density model of the tidal stream of NGC 3201 and
its H-R diagram from GDR2 after dust extinction correction.}}
\begin{center}
\begin{tabular}{rrrrrrrrrr}
\toprule

\multicolumn{1}{c}{N}
&\multicolumn{1}{c}{source\_id}
&\multicolumn{1}{c}{$\pi$}
&\multicolumn{1}{c}{$\delta$}
&\multicolumn{1}{c}{$\alpha$}
&\multicolumn{1}{c}{$\mu_\delta$}
&\multicolumn{1}{c}{$\mu_{\alpha*}$}
&\multicolumn{1}{c}{\scalebox{0.8}{$G_{\rm BP}\!-\!G_{\rm RP}$}}
&\multicolumn{1}{c}{$G$}
&\multicolumn{1}{c}{$\chi_{\rm sel}$}\\
&
&\multicolumn{1}{c}{\units{mas}}
&\multicolumn{1}{c}{\units{deg}}
&\multicolumn{1}{c}{\units{deg}}
&\multicolumn{1}{c}{\units{mas yr$^{-1}$}}
&\multicolumn{1}{c}{\units{mas yr$^{-1}$}}
&\multicolumn{1}{c}{\units{mag}}
&\multicolumn{1}{c}{\units{mag}}
&\multicolumn{1}{c}{\units{\scalebox{0.8}{yr$^{3}$ deg$^{-2}$ pc$^{-1}$ mas$^{-3}$}}}\\

\midrule

1&85111820717084288&$0.1651$&$+20.1781$&$42.2098$&$-7.9041$&$9.1652$&$0.8645$&$17.6426$&$8.1257\text{E}\:\!\:\!\minus\:\!02$\\
2&85463664437883264&$0.4112$&$+21.1340$&$42.5690$&$-7.1268$&$8.0718$&$1.1816$&$18.7109$&$5.3786\text{E}\:\!\:\!\minus\:\!02$\\
3&85259674966320512&$0.1913$&$+20.9929$&$42.8676$&$-7.1869$&$8.2032$&$1.2189$&$18.1123$&$2.9994\text{E}\:\!\:\!\minus\:\!01$\\
4&34258896831609984&$0.1405$&$+16.3693$&$45.1634$&$-9.6897$&$10.4887$&$0.9566$&$17.5377$&$5.3842\text{E}\:\!\:\!\minus\:\!02$\\
5&35721109857582464&$0.3439$&$+18.4065$&$45.9215$&$-9.1765$&$9.7779$&$1.0528$&$18.0348$&$6.4986\text{E}\:\!\:\!\minus\:\!02$\\
6&60185926474933632&$0.1295$&$+20.1024$&$46.0006$&$-7.6207$&$8.4273$&$0.8544$&$18.1997$&$5.3196\text{E}\:\!\:\!\minus\:\!02$\\
7&34901557082616576&$0.1467$&$+18.1273$&$46.4329$&$-9.8855$&$10.5074$&$0.7827$&$17.6309$&$6.2463\text{E}\:\!\:\!\minus\:\!02$\\
8&58640189220402432&$0.0641$&$+17.6070$&$47.2249$&$-9.7492$&$10.2046$&$0.8839$&$17.2243$&$2.0020\text{E}\:\!\:\!\minus\:\!01$\\
9&31456447850977536&$0.2736$&$+16.1339$&$47.4120$&$-9.7023$&$10.2854$&$1.1301$&$16.8443$&$2.1423\text{E}\:\!\:\!\minus\:\!01$\\
10&31185242140302080&$0.2541$&$+15.7169$&$48.4324$&$-10.1471$&$10.2482$&$0.9355$&$18.2953$&$8.7513\text{E}\:\!\:\!\minus\:\!02$\\
11&55510772969455232&$-0.0189$&$+17.0169$&$48.7138$&$-10.3461$&$10.9882$&$1.1011$&$16.5356$&$7.4714\text{E}\:\!\:\!\minus\:\!02$\\
12&54919579310541568&$0.2956$&$+16.7824$&$51.1901$&$-10.6029$&$10.7765$&$0.8089$&$17.7362$&$6.3037\text{E}\:\!\:\!\minus\:\!02$\\
13&37334471374125696&$0.1933$&$+11.8852$&$55.9878$&$-12.2495$&$12.5768$&$1.3554$&$17.3445$&$4.3200\text{E}\:\!\:\!\minus\:\!02$\\
14&36354359837010816&$0.2640$&$+10.6748$&$56.0746$&$-13.0978$&$13.3644$&$0.9893$&$18.5810$&$2.0527\text{E}\:\!\:\!\minus\:\!02$\\
15&3302763471907603840&$0.2579$&$+10.0905$&$57.2163$&$-15.0372$&$14.6607$&$0.9197$&$17.4677$&$2.7800\text{E}\:\!\:\!\minus\:\!02$\\
16&3302347405538192768&$0.1123$&$+9.0186$&$57.5283$&$-14.6794$&$13.5157$&$0.9866$&$18.1334$&$7.7986\text{E}\:\!\:\!\minus\:\!03$\\
17&3302517627978790784&$0.2305$&$+9.7887$&$58.3446$&$-13.4025$&$13.6600$&$0.8746$&$17.0933$&$1.1627\text{E}\:\!\:\!\minus\:\!02$\\
18&3273949498390088448&$0.2311$&$+6.2544$&$58.3769$&$-15.6878$&$15.0130$&$0.9952$&$17.1423$&$7.7895\text{E}\:\!\:\!\minus\:\!03$\\
19&3259376124600082688&$0.0439$&$+2.7632$&$63.5494$&$-18.0006$&$16.5966$&$1.0363$&$16.9342$&$1.0527\text{E}\:\!\:\!\minus\:\!02$\\
20&3283413643508855168&$0.4116$&$+3.2175$&$63.9090$&$-17.6916$&$16.5187$&$0.8983$&$18.2204$&$4.3555\text{E}\:\!\:\!\minus\:\!02$\\
21&3283707732806011776&$0.2904$&$+3.8795$&$66.0409$&$-16.9729$&$16.1823$&$1.0192$&$16.8527$&$7.6543\text{E}\:\!\:\!\minus\:\!03$\\
22&3278904202725707776&$0.4538$&$+0.6731$&$66.8839$&$-18.0428$&$16.9354$&$0.8860$&$18.3811$&$2.6335\text{E}\:\!\:\!\minus\:\!02$\\
23&3205031181848570240&$0.3451$&$-2.6066$&$69.8346$&$-19.8285$&$19.0347$&$0.6952$&$17.5299$&$6.7108\text{E}\:\!\:\!\minus\:\!03$\\
24&3229172192289896576&$0.5220$&$-1.8187$&$69.8390$&$-20.3993$&$19.0081$&$0.9991$&$19.1580$&$6.6485\text{E}\:\!\:\!\minus\:\!03$\\
25&3225714404316322048&$0.5186$&$-3.3168$&$72.0153$&$-20.3162$&$19.3406$&$0.9340$&$17.5908$&$1.3595\text{E}\:\!\:\!\minus\:\!02$\\
26&3225088713480594304&$0.1847$&$-3.0706$&$72.5461$&$-20.7489$&$19.8459$&$0.9562$&$15.6898$&$1.7710\text{E}\:\!\:\!\minus\:\!02$\\
27&3188312385993123968&$0.2624$&$-6.0464$&$72.9310$&$-23.3288$&$22.2830$&$0.7262$&$17.4658$&$7.0712\text{E}\:\!\:\!\minus\:\!03$\\
28&3225017378367519232&$0.1945$&$-3.3841$&$72.9781$&$-21.1877$&$20.1364$&$0.7722$&$18.1341$&$2.3440\text{E}\:\!\:\!\minus\:\!02$\\
29&3212479479773005696&$0.2639$&$-5.2492$&$73.6805$&$-22.6205$&$21.5271$&$0.6668$&$17.0226$&$2.9807\text{E}\:\!\:\!\minus\:\!02$\\
30&3187421678493781888&$0.3710$&$-6.0663$&$74.5026$&$-21.6876$&$20.4548$&$0.7595$&$17.7014$&$1.1591\text{E}\:\!\:\!\minus\:\!01$\\
31&3183889733612983296&$0.2107$&$-8.0801$&$75.1650$&$-24.2181$&$22.8004$&$0.9031$&$18.1802$&$5.9379\text{E}\:\!\:\!\minus\:\!03$\\
32&3182335646943552000&$0.3665$&$-9.8777$&$75.9863$&$-23.4187$&$22.8338$&$0.8467$&$17.0998$&$1.0542\text{E}\:\!\:\!\minus\:\!02$\\
33&3183733710337675136&$0.2786$&$-8.2173$&$76.5810$&$-22.5581$&$21.3874$&$0.8088$&$18.0632$&$6.5451\text{E}\:\!\:\!\minus\:\!02$\\
34&3182771152330545152&$0.0302$&$-9.2063$&$76.6356$&$-21.4523$&$20.5276$&$1.1081$&$18.9625$&$6.5732\text{E}\:\!\:\!\minus\:\!03$\\
35&2989805452906457216&$0.2265$&$-11.1401$&$77.4576$&$-24.2242$&$23.2763$&$0.9121$&$18.4459$&$5.6177\text{E}\:\!\:\!\minus\:\!03$\\
36&2985707611726244480&$0.2369$&$-13.1102$&$80.3573$&$-23.6520$&$23.4954$&$0.7765$&$17.4529$&$6.0301\text{E}\:\!\:\!\minus\:\!03$\\
37&2985448500643184896&$0.3116$&$-13.5979$&$81.3864$&$-22.5934$&$22.6676$&$0.8775$&$17.9810$&$1.6029\text{E}\:\!\:\!\minus\:\!02$\\
38&2985851682109593728&$0.3032$&$-12.5971$&$81.5529$&$-23.6375$&$23.2202$&$0.8874$&$16.3023$&$3.2902\text{E}\:\!\:\!\minus\:\!02$\\
39&2984359674895283072&$0.4909$&$-14.7883$&$83.5002$&$-22.2270$&$22.7071$&$0.8658$&$18.4725$&$6.4990\text{E}\:\!\:\!\minus\:\!03$\\
40&2971024729152562176&$0.1691$&$-16.9745$&$85.2735$&$-23.8125$&$23.9815$&$0.8599$&$17.9274$&$6.5887\text{E}\:\!\:\!\minus\:\!03$\\
41&2971248857726512128&$0.1681$&$-15.7976$&$85.7306$&$-23.1975$&$22.6792$&$1.0410$&$18.7166$&$9.8945\text{E}\:\!\:\!\minus\:\!03$\\
42&2967677373378985472&$0.4029$&$-18.1238$&$86.3326$&$-23.3190$&$24.2962$&$0.7562$&$17.2588$&$2.2519\text{E}\:\!\:\!\minus\:\!02$\\
43&2967743511577557120&$0.4519$&$-17.9255$&$87.4169$&$-22.9684$&$24.0202$&$0.8708$&$17.7389$&$3.1562\text{E}\:\!\:\!\minus\:\!02$\\
44&2967375007681446784&$0.2755$&$-18.6319$&$87.6077$&$-23.5623$&$24.2818$&$0.7710$&$17.5950$&$3.9916\text{E}\:\!\:\!\minus\:\!02$\\
45&2991860440139396608&$-0.0381$&$-16.8690$&$88.2245$&$-22.8467$&$22.9447$&$0.9589$&$18.7355$&$6.0765\text{E}\:\!\:\!\minus\:\!03$\\
46&2966606032438342144&$0.2674$&$-19.4013$&$88.5997$&$-22.2754$&$23.4095$&$0.8417$&$18.0282$&$1.4813\text{E}\:\!\:\!\minus\:\!01$\\
47&2966413828356833152&$0.2144$&$-19.5051$&$88.8025$&$-22.2084$&$23.3199$&$0.7694$&$17.3525$&$2.0891\text{E}\:\!\:\!\minus\:\!01$\\
48&2918181822364292736&$0.2471$&$-20.3285$&$89.1767$&$-22.3439$&$24.2402$&$0.8385$&$18.1034$&$1.6488\text{E}\:\!\:\!\minus\:\!02$\\
49&2990877854701797888&$0.3134$&$-17.4093$&$90.4209$&$-23.6439$&$23.5650$&$0.9256$&$17.1479$&$1.0600\text{E}\:\!\:\!\minus\:\!02$\\
50&2916935594654099712&$0.3217$&$-22.3985$&$90.8435$&$-22.0852$&$24.1436$&$0.7733$&$17.7879$&$5.6818\text{E}\:\!\:\!\minus\:\!03$\\
51&2917978004694480128&$0.2612$&$-20.5596$&$90.9743$&$-22.0869$&$23.9064$&$0.7235$&$17.1207$&$8.3623\text{E}\:\!\:\!\minus\:\!02$\\
52&2941157702670871680&$0.4038$&$-20.9649$&$91.8675$&$-21.8820$&$23.8108$&$0.7327$&$16.6851$&$5.0665\text{E}\:\!\:\!\minus\:\!02$\\
53&2941141931553193856&$0.3032$&$-21.1686$&$92.2365$&$-21.8213$&$23.1047$&$0.8605$&$17.9485$&$1.6438\text{E}\:\!\:\!\minus\:\!01$\\
54&2941295279065097472&$0.2902$&$-20.4353$&$92.5552$&$-22.8800$&$24.2738$&$0.6464$&$16.0479$&$2.0424\text{E}\:\!\:\!\minus\:\!02$\\
55&2937577417936027008&$0.2063$&$-22.4331$&$93.8184$&$-20.6428$&$22.4657$&$0.7654$&$17.5514$&$1.1616\text{E}\:\!\:\!\minus\:\!01$\\
56&2936748248729371008&$0.3291$&$-22.9566$&$95.0570$&$-21.4088$&$23.8275$&$0.7250$&$16.8219$&$2.2029\text{E}\:\!\:\!\minus\:\!02$\\
57&2937679290262547584&$0.0609$&$-21.8367$&$95.3396$&$-20.4362$&$22.4033$&$0.9023$&$18.3712$&$8.8452\text{E}\:\!\:\!\minus\:\!03$\\
58&2924361047651213184&$0.3045$&$-24.2864$&$96.6325$&$-20.8322$&$23.3102$&$0.6946$&$17.0423$&$1.1522\text{E}\:\!\:\!\minus\:\!01$\\
59&2923332351444848896&$0.2326$&$-25.4506$&$98.1197$&$-18.9226$&$21.7545$&$0.8182$&$16.1500$&$5.3253\text{E}\:\!\:\!\minus\:\!02$\\
60&2919762198531603456&$0.2841$&$-27.6668$&$98.8488$&$-18.8438$&$22.7615$&$0.9272$&$18.1750$&$5.9658\text{E}\:\!\:\!\minus\:\!03$\\
61&2923457523971828480&$0.3593$&$-25.2914$&$99.9176$&$-20.4000$&$23.7148$&$0.8881$&$18.0914$&$1.4250\text{E}\:\!\:\!\minus\:\!02$\\
62&2919619949213987200&$0.2580$&$-27.6074$&$100.5084$&$-19.2244$&$22.0807$&$0.9075$&$18.6624$&$2.2055\text{E}\:\!\:\!\minus\:\!02$\\
63&2918864520305483264&$0.4388$&$-27.7485$&$101.3058$&$-18.7304$&$22.2778$&$0.9241$&$18.1616$&$1.3436\text{E}\:\!\:\!\minus\:\!01$\\

\bottomrule

\end{tabular}
\end{center}

\label{sel0}
\end{table*}

\begin{table*}\addtocounter{table}{-1}
\caption[]{\small{\textit{- continued}}}
\begin{center}
\begin{tabular}{rrrrrrrrrr}
\toprule

\multicolumn{1}{c}{N}
&\multicolumn{1}{c}{source\_id}
&\multicolumn{1}{c}{$\pi$}
&\multicolumn{1}{c}{$\delta$}
&\multicolumn{1}{c}{$\alpha$}
&\multicolumn{1}{c}{$\mu_\delta$}
&\multicolumn{1}{c}{$\mu_{\alpha*}$}
&\multicolumn{1}{c}{\scalebox{0.8}{$G_{\rm BP}\!-\!G_{\rm RP}$}}
&\multicolumn{1}{c}{$G$}
&\multicolumn{1}{c}{$\chi_{\rm sel}$}\\
&
&\multicolumn{1}{c}{\units{mas}}
&\multicolumn{1}{c}{\units{deg}}
&\multicolumn{1}{c}{\units{deg}}
&\multicolumn{1}{c}{\units{mas yr$^{-1}$}}
&\multicolumn{1}{c}{\units{mas yr$^{-1}$}}
&\multicolumn{1}{c}{\units{mag}}
&\multicolumn{1}{c}{\units{mag}}
&\multicolumn{1}{c}{\units{\scalebox{0.8}{yr$^{3}$ deg$^{-2}$ pc$^{-1}$ mas$^{-3}$}}}\\

\midrule

64&2918802226100396032&$0.3037$&$-27.9478$&$101.8551$&$-18.4146$&$21.9682$&$1.1722$&$18.6582$&$8.6932\text{E}\:\!\:\!\minus\:\!02$\\
65&2919192204830525312&$0.1634$&$-27.5619$&$102.0146$&$-18.6500$&$22.3833$&$0.7739$&$16.7921$&$8.5157\text{E}\:\!\:\!\minus\:\!02$\\
66&5608974194743618304&$0.1431$&$-29.0134$&$103.0615$&$-18.4799$&$22.6030$&$0.9807$&$17.2841$&$6.2600\text{E}\:\!\:\!\minus\:\!03$\\
67&5604401257167040896&$0.1817$&$-31.3957$&$106.9045$&$-16.2835$&$21.7660$&$0.9967$&$18.2635$&$5.6973\text{E}\:\!\:\!\minus\:\!03$\\
68&5604387551929978624&$0.1216$&$-31.4825$&$107.4312$&$-15.7999$&$21.3128$&$0.8960$&$17.8501$&$2.8356\text{E}\:\!\:\!\minus\:\!02$\\
69&5604416031855199360&$0.2044$&$-31.2087$&$107.7262$&$-16.2320$&$21.3956$&$1.0559$&$17.4261$&$3.0635\text{E}\:\!\:\!\minus\:\!02$\\
70&5605860210317910656&$0.1027$&$-31.2627$&$108.2240$&$-16.1830$&$21.3612$&$0.8866$&$18.1165$&$5.1754\text{E}\:\!\:\!\minus\:\!03$\\
71&5605025887150553344&$0.2347$&$-31.7202$&$109.0093$&$-15.7407$&$21.4371$&$0.8631$&$17.5631$&$1.4133\text{E}\:\!\:\!\minus\:\!02$\\
72&5587654118824699136&$-0.0611$&$-35.6350$&$115.3624$&$-13.8094$&$19.3400$&$1.7499$&$20.5729$&$1.1787\text{E}\:\!\:\!\minus\:\!03$\\
73&5587324089239122688&$-0.2146$&$-36.4645$&$117.2742$&$-12.9198$&$19.9370$&$1.5662$&$19.2243$&$9.6739\text{E}\:\!\:\!\minus\:\!04$\\
74&5539044602383550720&$0.4356$&$-37.2963$&$119.2984$&$-11.0610$&$19.3346$&$2.4144$&$19.6161$&$5.4508\text{E}\:\!\:\!\minus\:\!04$\\
75&5424452710261263872&$-0.2191$&$-43.1483$&$140.7367$&$-4.6507$&$12.3840$&$1.6736$&$19.5148$&$1.1320\text{E}\:\!\:\!\minus\:\!03$\\
76&5424980029168882560&$0.3051$&$-43.1169$&$142.9833$&$-4.7781$&$11.2861$&$1.6641$&$19.0377$&$6.2820\text{E}\:\!\:\!\minus\:\!03$\\
77&5412717553938252544&$0.1385$&$-44.1282$&$144.6515$&$-4.4577$&$11.2894$&$1.1298$&$18.8000$&$1.2108\text{E}\:\!\:\!\minus\:\!02$\\
78&5412394400600530432&$-0.0005$&$-45.1641$&$146.4960$&$-3.4355$&$10.2152$&$1.0007$&$19.0347$&$6.6114\text{E}\:\!\:\!\minus\:\!03$\\
79&5411987340783459840&$0.0278$&$-44.7558$&$147.9285$&$-4.5372$&$9.6447$&$1.2994$&$19.8551$&$3.5455\text{E}\:\!\:\!\minus\:\!03$\\
80&5412063585038036096&$0.1458$&$-44.7961$&$148.5284$&$-2.7358$&$8.7963$&$1.0250$&$19.5906$&$4.3700\text{E}\:\!\:\!\minus\:\!03$\\
81&5411902609663942272&$0.2299$&$-44.7107$&$149.0004$&$-3.4370$&$10.5574$&$1.4347$&$20.2248$&$4.8709\text{E}\:\!\:\!\minus\:\!03$\\
82&5408678635415016576&$0.2071$&$-46.0086$&$149.5061$&$-2.6238$&$9.9105$&$0.9979$&$17.5241$&$1.2022\text{E}\:\!\:\!\minus\:\!01$\\
83&5411828465645835648&$0.1680$&$-44.9974$&$149.6303$&$-2.9013$&$9.7088$&$1.0017$&$19.0311$&$1.5012\text{E}\:\!\:\!\minus\:\!01$\\
84&5414808107804098432&$0.4901$&$-45.2014$&$150.4793$&$-3.6007$&$9.8601$&$1.1010$&$20.3807$&$6.1875\text{E}\:\!\:\!\minus\:\!03$\\
85&5408722242225130880&$-0.4469$&$-46.0245$&$150.6442$&$-3.1488$&$9.1825$&$1.1029$&$19.4458$&$2.9276\text{E}\:\!\:\!\minus\:\!03$\\
86&5414950318459476864&$0.0050$&$-44.4130$&$150.9149$&$-3.2737$&$8.4006$&$1.0093$&$19.3326$&$4.7786\text{E}\:\!\:\!\minus\:\!03$\\
87&5414797898662111744&$0.0673$&$-44.8816$&$151.1779$&$-3.7119$&$8.8205$&$1.1339$&$20.1646$&$5.2609\text{E}\:\!\:\!\minus\:\!03$\\
88&5414539822663808768&$0.3201$&$-45.6822$&$151.4893$&$-2.5543$&$9.8544$&$0.9911$&$18.8864$&$3.3081\text{E}\:\!\:\!\minus\:\!02$\\
89&5414564183718662016&$0.3195$&$-45.5630$&$151.5356$&$-2.8157$&$8.5250$&$1.0473$&$19.2840$&$4.9625\text{E}\:\!\:\!\minus\:\!02$\\
90&5414524021482764672&$0.5664$&$-46.0519$&$151.5567$&$-2.8607$&$9.5367$&$1.0047$&$20.1461$&$9.0886\text{E}\:\!\:\!\minus\:\!03$\\
91&5414568478687775616&$0.0000$&$-45.4720$&$151.7814$&$-1.9550$&$9.4562$&$1.0643$&$19.8742$&$1.9126\text{E}\:\!\:\!\minus\:\!02$\\
92&5414568203809849600&$0.0741$&$-45.4956$&$151.8199$&$-2.5132$&$9.2889$&$0.9280$&$18.2477$&$2.2365\text{E}\:\!\:\!\minus\:\!01$\\
93&5414470175475053440&$0.0939$&$-46.0117$&$152.0382$&$-2.0447$&$8.5806$&$0.9803$&$18.2909$&$1.5857\text{E}\:\!\:\!\minus\:\!01$\\
94&5407680210138297472&$0.3190$&$-47.0962$&$152.1243$&$-2.1293$&$8.9009$&$1.0210$&$19.3478$&$6.8538\text{E}\:\!\:\!\minus\:\!03$\\
95&5414514808777843968&$0.1885$&$-45.7093$&$152.1831$&$-3.2370$&$7.6466$&$1.1894$&$20.4158$&$3.7572\text{E}\:\!\:\!\minus\:\!03$\\
96&5414489799181004160&$0.5853$&$-46.0332$&$152.2404$&$-2.4198$&$9.1888$&$1.0093$&$19.8201$&$2.8798\text{E}\:\!\:\!\minus\:\!02$\\
97&5413703854532103040&$0.5427$&$-46.4378$&$152.2645$&$-1.4183$&$8.4917$&$1.2109$&$19.5160$&$1.0050\text{E}\:\!\:\!\minus\:\!02$\\
98&5413703403553024896&$0.4628$&$-46.5002$&$152.2753$&$-1.3476$&$8.8112$&$1.1792$&$19.9166$&$8.7105\text{E}\:\!\:\!\minus\:\!03$\\
99&5414501064881763584&$0.5115$&$-45.8547$&$152.2918$&$-2.1137$&$9.1032$&$0.9609$&$18.9478$&$7.2786\text{E}\:\!\:\!\minus\:\!02$\\
100&5413735804786642688&$0.5939$&$-46.2623$&$152.4463$&$-1.7846$&$8.6619$&$1.1320$&$18.7962$&$1.1387\text{E}\:\!\:\!\minus\:\!02$\\
101&5413736771158381056&$0.3367$&$-46.1567$&$152.4470$&$-2.6524$&$8.7161$&$1.0038$&$19.3687$&$5.5646\text{E}\:\!\:\!\minus\:\!02$\\
102&5413707531024130304&$0.1513$&$-46.4227$&$152.4619$&$-2.1157$&$8.7608$&$1.4619$&$14.2777$&$1.2925\text{E}\:\!\:\!\minus\:\!01$\\
103&5413686464201633024&$0.3246$&$-46.6598$&$152.4782$&$-1.3244$&$9.1873$&$1.0932$&$19.8740$&$8.2365\text{E}\:\!\:\!\minus\:\!03$\\
104&5413731922137962496&$0.0330$&$-46.3127$&$152.6005$&$-1.4314$&$8.4122$&$1.0919$&$19.1534$&$1.7983\text{E}\:\!\:\!\minus\:\!02$\\
105&5413746769842342912&$-0.2302$&$-45.9664$&$152.6060$&$-1.4081$&$8.1404$&$0.9907$&$19.7287$&$3.9730\text{E}\:\!\:\!\minus\:\!03$\\
106&5413694268165352704&$0.1283$&$-46.5443$&$152.6153$&$-1.9544$&$8.7215$&$1.2519$&$16.1443$&$1.2764\text{E}\:\!\:\!\plus\:\!00$\\
107&5413734224240710272&$-0.0087$&$-46.1877$&$152.6410$&$-2.3915$&$8.6046$&$1.0492$&$18.7688$&$1.4529\text{E}\:\!\:\!\minus\:\!01$\\
108&5413742268720312192&$0.1306$&$-46.0010$&$152.6414$&$-2.2343$&$8.8286$&$1.2550$&$15.9851$&$1.8786\text{E}\:\!\:\!\plus\:\!00$\\
109&5413721824672679552&$0.2962$&$-46.2819$&$152.6430$&$-2.0587$&$8.7421$&$1.0147$&$18.2996$&$5.4112\text{E}\:\!\:\!\minus\:\!01$\\
110&5413744364664262144&$0.0805$&$-45.9321$&$152.6470$&$-3.1119$&$9.2018$&$0.9047$&$18.5161$&$9.4405\text{E}\:\!\:\!\minus\:\!03$\\
111&5413743295212382208&$0.3754$&$-45.9628$&$152.7251$&$-2.6614$&$8.3346$&$1.0136$&$18.5626$&$4.1028\text{E}\:\!\:\!\minus\:\!02$\\
112&5413837204676171136&$0.1280$&$-45.9449$&$152.7622$&$-2.3460$&$8.2183$&$1.0011$&$17.5443$&$1.7796\text{E}\:\!\:\!\minus\:\!02$\\
113&5413720622081853824&$0.2782$&$-46.3067$&$152.7667$&$-1.9144$&$8.6098$&$0.9940$&$17.6895$&$3.9211\text{E}\:\!\:\!\minus\:\!01$\\
114&5413742951614948096&$-0.5966$&$-46.0033$&$152.7743$&$-1.8458$&$8.8833$&$1.0139$&$19.5849$&$3.3428\text{E}\:\!\:\!\minus\:\!03$\\
115&5413827996266328064&$-0.0091$&$-45.8243$&$152.9341$&$-2.2238$&$8.5120$&$0.9132$&$17.3248$&$6.2819\text{E}\:\!\:\!\minus\:\!02$\\
116&5413634654017870336&$0.2101$&$-46.9412$&$152.9942$&$-2.0045$&$8.5502$&$1.2115$&$16.4645$&$1.9360\text{E}\:\!\:\!\minus\:\!01$\\
117&5413835894705134208&$-0.2688$&$-45.7065$&$153.0241$&$-1.8431$&$8.3532$&$0.9813$&$20.1126$&$5.5075\text{E}\:\!\:\!\minus\:\!03$\\
118&5414389537468984064&$0.1168$&$-44.3112$&$153.8987$&$-3.5476$&$8.3693$&$1.0589$&$18.4613$&$3.8191\text{E}\:\!\:\!\minus\:\!03$\\
119&5414057892972812160&$-0.0016$&$-45.2555$&$155.4083$&$-2.1204$&$7.7283$&$0.8449$&$17.6746$&$4.6860\text{E}\:\!\:\!\minus\:\!03$\\
120&5365783216481459840&$0.1342$&$-46.8392$&$155.8449$&$-1.4280$&$8.0469$&$1.1206$&$19.4101$&$1.7484\text{E}\:\!\:\!\minus\:\!02$\\
121&5365391073082942336&$0.2777$&$-47.3759$&$155.8568$&$-1.4592$&$8.4773$&$0.8594$&$17.7940$&$1.0584\text{E}\:\!\:\!\minus\:\!02$\\
122&5365780398982621696&$0.2621$&$-46.9529$&$155.8643$&$-1.3769$&$7.8673$&$1.0697$&$19.2203$&$2.2998\text{E}\:\!\:\!\minus\:\!02$\\
123&5365399383845390976&$-0.0230$&$-47.2051$&$155.9170$&$-1.2253$&$7.5164$&$0.9504$&$18.4153$&$1.0633\text{E}\:\!\:\!\minus\:\!02$\\
124&5365789504313726080&$0.0047$&$-46.7820$&$155.9225$&$-1.7586$&$6.9475$&$1.0258$&$19.1421$&$3.0141\text{E}\:\!\:\!\minus\:\!03$\\
125&5365876365730787328&$0.3279$&$-46.2400$&$155.9265$&$-2.0997$&$8.1445$&$0.9316$&$18.5095$&$5.3858\text{E}\:\!\:\!\minus\:\!02$\\
126&5365792016874464640&$0.0931$&$-46.6439$&$155.9438$&$-1.7115$&$8.1138$&$0.9518$&$17.1936$&$1.0490\text{E}\:\!\:\!\minus\:\!01$\\

\bottomrule
\end{tabular}
\end{center}

\label{sel1}
\end{table*}

\begin{table*}\addtocounter{table}{-1}
\caption[]{\small{\textit{- continued}}}
\begin{center}
\begin{tabular}{rrrrrrrrrr}
\toprule

\multicolumn{1}{c}{N}
&\multicolumn{1}{c}{source\_id}
&\multicolumn{1}{c}{$\pi$}
&\multicolumn{1}{c}{$\delta$}
&\multicolumn{1}{c}{$\alpha$}
&\multicolumn{1}{c}{$\mu_\delta$}
&\multicolumn{1}{c}{$\mu_{\alpha*}$}
&\multicolumn{1}{c}{\scalebox{0.8}{$G_{\rm BP}\!-\!G_{\rm RP}$}}
&\multicolumn{1}{c}{$G$}
&\multicolumn{1}{c}{$\chi_{\rm sel}$}\\
&
&\multicolumn{1}{c}{\units{mas}}
&\multicolumn{1}{c}{\units{deg}}
&\multicolumn{1}{c}{\units{deg}}
&\multicolumn{1}{c}{\units{mas yr$^{-1}$}}
&\multicolumn{1}{c}{\units{mas yr$^{-1}$}}
&\multicolumn{1}{c}{\units{mag}}
&\multicolumn{1}{c}{\units{mag}}
&\multicolumn{1}{c}{\units{\scalebox{0.8}{yr$^{3}$ deg$^{-2}$ pc$^{-1}$ mas$^{-3}$}}}\\

\midrule

127&5365873445153019008&$0.0975$&$-46.2453$&$155.9449$&$-1.3040$&$7.6397$&$1.3341$&$19.7425$&$5.1718\text{E}\:\!\:\!\minus\:\!03$\\
128&5365865959033473152&$0.2407$&$-46.3788$&$156.0361$&$-2.1501$&$8.4814$&$0.9965$&$19.1296$&$9.3859\text{E}\:\!\:\!\minus\:\!03$\\
129&5365785965260649216&$0.1707$&$-46.7892$&$156.0661$&$-1.4949$&$7.4019$&$0.9192$&$18.7294$&$1.9158\text{E}\:\!\:\!\minus\:\!02$\\
130&5365819019327251840&$0.5079$&$-46.3904$&$156.0674$&$-1.4804$&$7.4436$&$1.0385$&$19.3859$&$5.7243\text{E}\:\!\:\!\minus\:\!03$\\
131&5365784178553932544&$-0.1314$&$-46.9102$&$156.0887$&$-1.6961$&$7.5548$&$1.1499$&$19.4999$&$7.6562\text{E}\:\!\:\!\minus\:\!03$\\
132&5365817163903821184&$0.0751$&$-46.5291$&$156.1047$&$-1.6714$&$8.3504$&$0.9041$&$18.6001$&$1.5036\text{E}\:\!\:\!\minus\:\!02$\\
133&5365825414540727296&$0.1149$&$-46.3280$&$156.1078$&$-2.0940$&$7.8422$&$1.1164$&$19.7862$&$9.0039\text{E}\:\!\:\!\minus\:\!03$\\
134&5365827476119277184&$-0.0518$&$-46.2354$&$156.2146$&$-1.9832$&$7.0940$&$1.0906$&$19.5436$&$3.1556\text{E}\:\!\:\!\minus\:\!03$\\
135&5365810261890643840&$0.2137$&$-46.7570$&$156.2188$&$-2.0729$&$6.9264$&$0.7268$&$18.8097$&$2.9639\text{E}\:\!\:\!\minus\:\!03$\\
136&5365921445707194624&$0.2320$&$-46.1857$&$156.2351$&$-1.5871$&$7.4213$&$1.0728$&$19.0933$&$1.1313\text{E}\:\!\:\!\minus\:\!02$\\
137&5365193706449182592&$0.3000$&$-47.4672$&$156.2907$&$-1.2538$&$7.5130$&$0.8539$&$18.2828$&$2.5675\text{E}\:\!\:\!\minus\:\!02$\\
138&5365768338714318080&$0.7031$&$-47.0099$&$156.3480$&$-1.1768$&$8.0557$&$1.1333$&$19.4720$&$3.4492\text{E}\:\!\:\!\minus\:\!03$\\
139&5365819775241396224&$0.0269$&$-46.4617$&$156.3699$&$-2.0587$&$8.1793$&$0.9681$&$18.6001$&$1.4942\text{E}\:\!\:\!\minus\:\!02$\\
140&5365190682792082048&$0.1500$&$-47.5861$&$156.3725$&$-1.4167$&$8.1328$&$0.8341$&$18.3808$&$2.2653\text{E}\:\!\:\!\minus\:\!02$\\
141&5365809437262578176&$0.2534$&$-46.4901$&$156.4128$&$-1.7321$&$7.5130$&$1.0855$&$19.4212$&$8.7985\text{E}\:\!\:\!\minus\:\!03$\\
142&5365578294999961344&$0.4987$&$-47.1869$&$156.4923$&$-1.9379$&$7.6787$&$0.8898$&$18.8358$&$5.5409\text{E}\:\!\:\!\minus\:\!03$\\
143&5365579016555060096&$0.2704$&$-47.1131$&$156.5143$&$-1.9269$&$7.0518$&$1.0797$&$19.3789$&$4.2389\text{E}\:\!\:\!\minus\:\!03$\\
144&5365576065920333440&$0.1539$&$-47.1961$&$156.6824$&$-1.6095$&$7.6348$&$1.5148$&$12.9366$&$1.0524\text{E}\:\!\:\!\plus\:\!01$\\
145&5365898355962610688&$0.3840$&$-46.5396$&$156.7343$&$-1.5688$&$7.0853$&$0.9684$&$18.9476$&$5.4961\text{E}\:\!\:\!\minus\:\!03$\\
146&5365601629562571008&$0.2543$&$-47.0240$&$156.8280$&$-0.9371$&$7.3243$&$1.1609$&$18.9579$&$5.2161\text{E}\:\!\:\!\minus\:\!03$\\
147&5365736178000020864&$0.3027$&$-46.1942$&$157.4884$&$-2.2346$&$7.7748$&$0.9663$&$17.0385$&$2.0367\text{E}\:\!\:\!\minus\:\!02$\\
148&5364616737729276928&$0.3602$&$-47.7895$&$159.6636$&$-0.6013$&$7.2650$&$1.0342$&$19.2178$&$4.9192\text{E}\:\!\:\!\minus\:\!03$\\
149&5366070189016080384&$0.4429$&$-47.5857$&$160.3078$&$-1.1778$&$6.6145$&$1.1561$&$18.3634$&$7.2854\text{E}\:\!\:\!\minus\:\!03$\\
150&5366058407925709568&$-0.0212$&$-47.7929$&$160.3254$&$-1.1813$&$6.5427$&$0.9288$&$17.8935$&$7.3902\text{E}\:\!\:\!\minus\:\!03$\\
151&5363031246257881984&$0.4249$&$-48.2777$&$161.0511$&$-0.5582$&$6.8420$&$0.9608$&$18.1045$&$9.4285\text{E}\:\!\:\!\minus\:\!03$\\
152&5363251045501068160&$0.3074$&$-47.6732$&$163.6266$&$-1.0451$&$6.0923$&$0.9211$&$17.9304$&$2.4429\text{E}\:\!\:\!\minus\:\!02$\\
153&5374364767295481856&$0.1341$&$-47.8018$&$168.1795$&$-0.4224$&$4.5801$&$0.7357$&$18.7395$&$1.1996\text{E}\:\!\:\!\minus\:\!02$\\
154&5374570105393379072&$0.1369$&$-47.2856$&$169.4423$&$-0.8148$&$4.3892$&$1.2607$&$14.4835$&$1.2424\text{E}\:\!\:\!\minus\:\!01$\\
155&5373850234514547712&$0.2316$&$-47.6973$&$171.1533$&$-0.3389$&$3.6895$&$1.0015$&$19.0083$&$1.0363\text{E}\:\!\:\!\minus\:\!02$\\
156&5373847481435744512&$0.1238$&$-47.1102$&$171.3746$&$-1.1421$&$3.6006$&$1.0540$&$18.0519$&$9.0333\text{E}\:\!\:\!\minus\:\!03$\\
157&5373628957800529280&$0.1411$&$-47.6738$&$172.1138$&$-0.7113$&$4.0836$&$0.8235$&$18.1147$&$2.4378\text{E}\:\!\:\!\minus\:\!02$\\
158&5375101959778889216&$-0.2126$&$-47.3622$&$173.2150$&$-0.5569$&$3.3159$&$0.8618$&$19.1182$&$9.8719\text{E}\:\!\:\!\minus\:\!03$\\
159&5375091376982888192&$0.2061$&$-47.4674$&$173.5502$&$-0.5459$&$3.5834$&$0.8895$&$18.6912$&$4.3155\text{E}\:\!\:\!\minus\:\!02$\\
160&5372062218149409536&$0.1720$&$-47.9061$&$174.1886$&$-0.4939$&$3.0765$&$1.0942$&$19.4563$&$7.8447\text{E}\:\!\:\!\minus\:\!03$\\
161&5372147915636764544&$0.5759$&$-47.4353$&$174.6326$&$-0.5447$&$3.6150$&$0.7032$&$19.3034$&$9.9653\text{E}\:\!\:\!\minus\:\!03$\\
162&5371948693572713856&$-0.0380$&$-47.4375$&$175.3823$&$-1.2254$&$2.8094$&$0.9519$&$19.9073$&$7.7881\text{E}\:\!\:\!\minus\:\!03$\\
163&5372345617270244480&$0.0380$&$-46.9986$&$175.7310$&$-1.5697$&$3.1700$&$1.2063$&$20.0780$&$8.1299\text{E}\:\!\:\!\minus\:\!03$\\
164&5372254082932197248&$0.2239$&$-47.7630$&$176.1961$&$-0.4725$&$3.3897$&$1.0882$&$19.2860$&$2.2251\text{E}\:\!\:\!\minus\:\!02$\\
165&5372296551563595520&$-0.2300$&$-47.1799$&$176.3097$&$-0.4701$&$3.3724$&$0.9962$&$19.5314$&$2.4790\text{E}\:\!\:\!\minus\:\!02$\\
166&5372386608438137600&$-0.2060$&$-46.9739$&$176.5707$&$-1.1318$&$3.1125$&$1.0581$&$19.7879$&$4.4448\text{E}\:\!\:\!\minus\:\!02$\\
167&5372378877496792064&$-0.0659$&$-47.1248$&$176.6575$&$-1.2263$&$3.3820$&$0.7079$&$19.4287$&$1.3672\text{E}\:\!\:\!\minus\:\!02$\\
168&5371655196988321152&$0.7329$&$-46.8830$&$177.0731$&$-0.5633$&$3.7295$&$0.7188$&$20.2820$&$1.4290\text{E}\:\!\:\!\minus\:\!02$\\
169&5377605302940159232&$0.5001$&$-46.8489$&$177.7541$&$-0.1653$&$3.7658$&$0.9804$&$20.0362$&$1.3828\text{E}\:\!\:\!\minus\:\!02$\\
170&5377724707328398592&$0.3648$&$-46.4346$&$178.2477$&$-1.0006$&$3.7925$&$0.8342$&$20.0085$&$8.0794\text{E}\:\!\:\!\minus\:\!03$\\

\bottomrule

\end{tabular}
\end{center}

\label{sel2}
\end{table*}


\section{Definition of stream coordinates}\label{App3}

We have defined stream spherical coordinates on the sky by defining
the angle $\phi_1$ along a major circle that approximately contains
the stream, and $\phi_2$ to be the polar angle from the axis
perpendicular to this major circle. An approximate adjustment to
these coordinates by eye has resulted in the following coordinate
transformation matrix from the usual equatorial coordinates
($\alpha, \delta$):
\begin{multline}
\begin{pmatrix}
\cos\var{\phi_2}\cos\var{\phi_1}\\
\cos\var{\phi_2}\sin\var{\phi_2}\\
\sin\var{\phi_2}
\end{pmatrix}
= M
\times
\begin{pmatrix}
\cos\var{\delta}\cos\var{\alpha}\\
\cos\var{\delta}\sin\var{\alpha}\\
\sin\var{\delta}
\end{pmatrix} ~,
\end{multline}
where the transformation matrix is:
\begin{equation}
M = 
\begin{pmatrix}
-0.6209 &  0.2992 & -0.7245 \\
-0.4004 & -0.9157 & -0.0350 \\
-0.6739 &  0.2684 &  0.6883 \\
\end{pmatrix} ~.
\end{equation}


\bsp	
\label{lastpage}

\end{document}